\documentclass[twocolumn]{aastex63}
\usepackage{times}
\usepackage{amsmath}
\usepackage{textcomp}
\usepackage{amssymb}
\usepackage{natbib}
\usepackage{float}
\usepackage{color}
\usepackage{graphicx}
\usepackage{hyperref}
\newcommand{\msun}{\ensuremath{M_{\odot}}}
\newcommand{\lum}{erg\,s$^{-1}$}

\newcommand{\xmm}{{\it XMM-Newton}}
\newcommand{\swift}{{\it Swift}}
\newcommand{\chandra}{{\it Chandra}}

\newcommand{\ergflux}{\mbox{${\rm \, erg \,\, cm^{-2} \, s^{-1}}$}}
\newcommand{\gm}{$\gamma$}
\newcommand{\mbh}{$M_{\rm BH}$}

\newcommand{\km}{\mbox{${\rm \, km \,\,  s^{-1}}$}}

\newcommand{\hbeta}{H{$\beta$}}

\newcommand{\FeII}{Fe{\sevenrm II}}

\newcommand{\OIII}{[O{\sevenrm\,III}]}
\newcommand{\OIIIa}{[O{\sevenrm\,III}]\,$\lambda$4959}
\newcommand{\OIIIb}{[O{\sevenrm\,III}]\,$\lambda$5007}
\newcommand{\OIIIab}{[O{\sevenrm\,III}]\,$\lambda\lambda$4959,5007}

 \font\sevenrm=cmr7 scaled 1000

\received{\today}
\revised{\today}
\accepted{\today}
\submitjournal{ApJ}

\shorttitle{Narrow-Line Seyfert 1 Galaxies in DESI DR1}
\shortauthors{Paliya et al.}

\begin{document}
\title{Narrow-Line Seyfert 1 Galaxies in the Dark Energy Spectroscopic Instrument Data Release 1}

\correspondingauthor{Vaidehi S. Paliya}
\email{vaidehi.s.paliya@gmail.com}

\author[0000-0001-7774-5308]{Vaidehi S. Paliya}
\affiliation{Inter-University Centre for Astronomy and Astrophysics (IUCAA), SPPU Campus, Pune 411007, India}

\author[0000-0002-8377-9667]{Suvendu Rakshit}
\affiliation{Aryabhatta Research Institute of Observational Sciences, Nainital–263001, Uttarakhand, India}

\author[0000-0002-3433-4610]{Alberto Dom{\'{\i}}nguez}
\affiliation{IPARCOS and Department of EMFTEL, Universidad Complutense de Madrid, E-28040 Madrid, Spain}

\author[0000-0002-4464-8023]{D. J. Saikia}
\affiliation{Fakultat f\"ur Physik, Universit\"at Bielefeld, Postfach 100131, D-33501 Bielefeld, Germany}
\affiliation{Assam Don Bosco University, Guwahati 781017, Assam, India}

\author[0000-0002-4998-1861]{C. S. Stalin}
\affiliation{Indian Institute of Astrophysics, Bengaluru, 560034, India}

\begin{abstract}
Narrow-line Seyfert 1 (NLSy1) galaxies are peculiar active galactic nuclei (AGN) known to exhibit a variety of intriguing observational features from low-frequency radio waves to high-energy \gm~rays. As of now, NLSy1 catalogs are primarily based on optical spectroscopic observations from the Sloan Digital Sky Survey (SDSS). Here we report, for the first time, a new catalog of NLSy1 galaxies using the high-quality optical spectroscopic observations made public in the first data release of the Dark Energy Spectroscopic Instrument (DESI). We performed a detailed spectral decomposition of more than 71,000 optical spectra of AGN not included in the SDSS catalog and located at $z<0.9$. From this sample, we identify 18749 objects as NLSy1 galaxies for the first time. We also supplement the NLSy1 catalog with a sample of broad-line Seyfert 1 galaxies. The NLSy1 galaxies identified in the DESI data tend to have slightly higher bolometric luminosities and lower black hole masses (though with large dispersions), leading to the higher Eddington ratios than those of the SDSS-NLSy1 sample matched in redshifts and absolute $B$-band magnitudes. Moreover, the fraction of DESI-NLSy1 galaxies detected in the radio, X-ray, and \gm-ray catalogs was found to be lower than that of SDSS-NLSy1 sources. We conclude that deeper multiwavelength investigations of these enigmatic AGN will help unravel the low-luminosity end of the NLSy1 population. The catalog has been made available at \url{https://www.ucm.es/blazars/seyfert} and Zenodo (\url{https://doi.org/10.5281/zenodo.20484681}).

\end{abstract}

\keywords{methods: data analysis --- galaxies: active --- galaxies: Seyfert --- techniques: imaging spectroscopy}

\section{Introduction}

In the active galactic nuclei (AGN) zoo, narrow-line Seyfert 1 (NLSy1) galaxies are one of the most peculiar members. They were classified based on the narrowness of the broad \hbeta~emission line (full width at half maximum or FWHM$<$2000 \km), the strength of the forbidden \OIII~line with respect to \hbeta~(\OIIIb/\hbeta~flux ratio $<$3) and identified as low-luminosity AGN \citep[$B$-band absolute magnitude $M_{\rm B}>-22.25$;][]{1985ApJ...297..166O,1989ApJ...342..224G,2010A&A...518A..10V}. The observations of rapid X-ray variability, steep X-ray spectra, strong outflows, and strong \FeII~complexes from some of these objects indicated NLSy1 galaxies to be powered by low-mass black holes ($M_{\rm BH}\sim10^{6-8}$ \msun) with a high accretion rate \citep[e.g.,][]{1996A&A...305...53B,1999ApJS..125..297L,1999ApJS..125..317L,2004ApJ...606L..41G,2010ApJS..187...64G}. Strong outflows allows us to study the feedback process in AGN \citep[e.g.,][]{2010MNRAS.401....7H,2012ARA&A..50..455F,2016MNRAS.458..816H}. They are usually detected as flux-excess blueshifted with respect to the core component of \OIIIb~emission line, leading to the observation of an asymmetric line profile \citep[cf.][]{2008ApJ...680..926K}. Strong outflows, with widths varying from a few hundred to a few thousand \km~are often detected in AGN with high Eddington rate such as NLSy1 galaxies where rapid accretion near the Eddington limit can trigger radiation-pressure driven outflows \citep[][]{2002ApJ...565...78B,2004AJ....127.1799G,2012AJ....143...83X,2018MNRAS.477.5115K,2018ApJ...865....5R}. 

With the availability of high-quality observations delivered by current generation of telescopes, it became evident that NLSy1 galaxies are intriguing sources. In the radio band, contrary to $\sim10-15\%$ of the general AGN population being radio loud, only $\sim$3-5\% NLSy1s were radio detected of which $\sim60-80\%$ were found to be radio-loud in different samples \citep[e.g.,][]{2018MNRAS.480.1796S,2024MNRAS.527.7055P}. Superluminal motion has been observed from several objects of this class indicating relativistic beaming and identification of large-scale ($>$100 kiloparsec) radio jets has also been reported \citep[e.g.,][]{2016RAA....16..176F,2018ApJ...869..173R,2024Galax..12....8K,2024ApJ...963...32C,2025ApJ...995..125U}. Furthermore, NLSy1 objects show flux variations at infrared (IR) and optical bands including nightly brightness changes \citep[][]{2013MNRAS.428.2450P,2017ApJ...842...96R,2017MNRAS.466.2679K,2019MNRAS.483.2362R,2025MNRAS.543..121S}. Recently, using Zwicky Transient Facility observations, \citet[][]{2021ApJ...920...56F} reported NLSy1 galaxies to exhibit rapidly rising and long-lasting flaring episodes in the optical band. Moreover, the high-resolution host galaxy imaging of some radio-loud NLSy1 galaxies has revealed traces of mergers and interactions, thus indicating a possible connection between galaxy mergers and triggering of jets \citep[][]{2016MNRAS.460.3202O,2019AJ....157...48B,2020ApJ...892..133P}. These sources are typically hosted in disk or spiral galaxies \citep[e.g.,][]{2003AJ....126.1690C}.

The X-ray observations of NLSy1 galaxies led to the detection of interesting spectral features, such as strong, often variable, soft X-ray excess, distorted Fe-K$\alpha$ emission line, and reflection dominated hard X-ray emission. These findings suggested complex radiative environment in the immediate vicinity of the central supermassive black hole where the interaction between matter and energy takes place \citep[e.g.,][]{2009Natur.459..540F,2014MNRAS.443.1723P,2015MNRAS.451.4375F,2017MNRAS.468.3489K,2018MNRAS.477.3247B,2019MNRAS.490..683P,2020MNRAS.496.2922M}. Variable \gm-ray emission has also been detected from a few radio-loud NLSy1 galaxies by the Fermi Large Area Telescope \citep[LAT; cf.][]{2009ApJ...707L.142A,2011MNRAS.413.2365C,2015AJ....149...41P,2018ApJ...853L...2P}. This observation provides unambiguous evidence for the presence of closely aligned relativistic jets in these objects similar to blazar class of AGN. Considering the fact that NLSy1 galaxies are thought to be low-luminosity AGN in the early phase of their evolution \citep[][]{2000MNRAS.314L..17M}, the \gm-ray detected objects of this class have been proposed as nascent blazars \citep[see, e.g.,][]{2015ANA...575A..13F,2019ApJ...872..169P,2020ApJ...892..133P}. 

As of now, the samples of NLSy1 galaxies were prepared using the various data releases of the Sloan Digital Sky Survey (SDSS). For example, using the SDSS early data release (DR) and DR3, \citet[][]{2002AJ....124.3042W} and \citet[][]{2006ApJS..166..128Z} reported the identification of 150 and 2011 NLSy1s, respectively. \citet[][]{2017ApJS..229...39R} analyzed optical spectra released in SDSS-DR12 and reported a catalog of 11101 NLSy1 galaxies. More recently, a catalog of 22656 NLSy1 objects was published using SDSS-DR17 \citep[][]{2024MNRAS.527.7055P}.

The Dark Energy Spectroscopic Instrument (DESI) collaboration recently announced the first release of the optical spectroscopic data acquired during the first 13 months of the DESI main survey \citep[hereafter DESI-DR1;][]{2025arXiv250314745D}. With over 18 million optical spectra released in DESI-DR1, this is the largest ever optical spectroscopic survey ever conducted surpassing SDSS. The optical spectroscopy is being carried out using DESI mounted at Mayall 4-m telescope at Kitt Peak National Observatory, Arizona, United States of America. The details of the DESI-DR1 can be found in \citet[][]{2025arXiv250314745D}. Considering the broader sky coverage and deeper sensitivity of DESI compared to SDSS, it is imperative to increase the sample size of NLSy1 galaxies using DESI-DR1 dataset. Also considering the ongoing and upcoming wide-field multiwavelength surveys, e.g., LOw Frequency ARray (LOFAR) and eROSITA, such an enlarged sample of NLSy1 galaxies will enable us to carry out detailed investigations of some of the faintest members of this class of AGN. With this objective in mind, we carried out a systematic spectroscopic decomposition analysis of a large sample of DESI-DR1 AGN optical spectra, and report here, for the first time, the identification of 18749 NLSy1 galaxies.

In Section~\ref{sec2}, we briefly describe the steps adopted to make the parent sample. The data analysis methodology is elaborated in Section~\ref{sec3}. We present the new NLSy1 catalog in Section~\ref{sec4} and discuss their multiwavelength properties in Section~\ref{sec5}. The results are summarized in Section~\ref{sec6}. Throughout, a flat cosmology with $H_0 = 70~{\rm km~s^{-1}~Mpc^{-1}}$ and $\Omega_{\rm M} = 0.3$ was used.

\section{Sample Selection}\label{sec2}
We considered the AGN/QSO value added catalog (VAC) of 17995599 objects\footnote{\url{https://data.desi.lbl.gov/doc/releases/dr1/vac/agnqso/}} (S. Juneau et al., in preparation). Using the emission line measurements from FastSpecFit\footnote{\url{https://fastspecfit.readthedocs.io/en/latest/iron.html}}, the catalog includes AGN identified using optical and ultraviolet diagnostic diagrams and mid-infrared classifications based on Tractor WISE photometry \citep[][]{1987ApJS...63..295V,2001ApJ...556..121K,2016ascl.soft04008L}. To optimize the sample selection, we adopted the following criteria:

\begin{enumerate}
    \item Only considered the sources that have a primary coadded spectrum, i.e, the keyword {\tt zcat\_primary} is set to be {\tt True}.
    \item Only selected objects that have no measured redshift warning, i.e, {\tt zwarn = 0}.
    \item We cross-matched DESI sources with SDSS-DR17 using a search radius of 3 arcseconds, and removed all the matches. This step ensured the selection of sources whose optical spectra were not analyzed earlier for identifying NLSy1 galaxies.
    \item The AGN/QSO VAC provides spectral parameters of several emission lines including \hbeta~broad component. We considered only those sources that have a positive \hbeta~flux (keyword {\tt HBETA\_BROAD\_FLUX} $>$0), the ratio of the \hbeta~flux to the corresponding uncertainty $>$3, and the dispersion of the \hbeta~broad component $>$300. The last condition provided an initial filtering of objects in which the broad \hbeta~component is either weak or undetected.
    \item We manually inspected the optical spectra of 104 sources that have a positive {\tt HBETA\_BROAD\_FLUX} but the quoted redshift was $z>1.29$. We corrected the redshifts of 45 such sources and included them in the sample. The \hbeta~line was not covered in the optical spectra of the remaining sources.
    \item Since NLSy1 classification also requires \OIIIb~emission line flux values, we applied a redshift cutoff of $z\leq0.9$ to ensure that \OIIIb~lines falls into the wavelength range covered by DESI (3600 \AA$-$9800 \AA). Additionally, we applied a lower redshift limit of $z\geq0.01$ to avoid contamination from Galactic objects. 
\end{enumerate}

Following the above-mentioned steps, we were left with 71918 objects. The optical spectra of these sources were then analyzed following the methodology described in the next section.

\section{Optical Spectroscopic Data Analysis}\label{sec3}
We followed the commonly adopted approach to decompose the optical spectrum which involves fitting and subtracting the continuum radiation and modeling of the emission lines \citep[e.g.,][]{2017ApJS..229...39R,2023ApJS..267...19I}. For spectral analysis, we considered the rest-frame wavelength range of 3800$-$5500 \AA, which contains the emission lines of interest, i.e., \hbeta~and \OIIIab, and is large enough to reliably measure the continuum. 

In the first step, the spectrum was brought to the rest-frame and corrected for Galactic reddening following \citet[][]{1998ApJ...500..525S}. Next, we modeled the continuum emission with a combination of a power-law and a composite host-galaxy template. Since NLSy1 galaxies usually exhibit strong \FeII~emission, we also used the optical \FeII~template of \citet[][]{1992ApJS...80..109B}. 

To model the host galaxy emission, we adopted the prior-informed AGN host spectral decomposition technique described in \citet[][]{2024ApJ...974..153R} and implemented in publicly available multicomponent spectral fitting code {\tt PyQSOFit}\footnote{\url{https://github.com/legolason/PyQSOFit}} \citep[][]{2018ascl.soft09008G}. In particular, we adopted galaxy and quasar templates from principal component analysis (PCA) developed by \citet[][]{2004AJ....128..585Y,2004AJ....128.2603Y}. The considered assumption is that the observed spectrum is a combination of two sets of eigenspectra from pure galaxy and quasar samples. Therefore, integrating the PCA spectra of both components enables us to separate them \citep[e.g.,][]{2006AJ....131...84V}. However, PCA spectra from galaxy and quasar template libraries can be degenerate. More importantly, for the decomposition of low signal-to-noise ratio spectra, high-order PCA templates might overfit noise, thus leading to their disproportionate weighting in the spectral fitting. In this regard, the use of the PCA eigen coefficient distribution as prior information reduces the degeneracy of PCA templates between quasar and galaxy templates, thus significantly avoiding the overfitting issue \citep[see,][for details]{2024ApJ...974..153R}.

Several previous studies have found that NLSy1 galaxies typically exhibit broad Balmer lines with extended wings which are better represented with a Lorentzian profile \citep[cf.][]{1996ApJS..106..341M,2001A&A...372..730V,2002ApJ...566L..71S,2012MNRAS.426.3086G}. Therefore, the broad \hbeta~component was modeled with a Lorentzian profile and the narrow component was reproduced with a Gaussian shape. Furthermore, strong outflows have been detected in several NLSy1 galaxies, thus we model the \OIII~doublet with double Gaussian functions, one each for the core and wing. We also applied the constraints that the FWHM of the wing should be larger than that of the core component and that the peak of the core should be larger than the wing component peak. The widths of the narrow \hbeta~and \OIIIa~were tied to that of \OIIIb~emission line. The flux ratio of \OIII~doublet was fixed to the theoretical value of 3 \citep[e.g.,][]{2011ApJS..194...45S}. The constraint that the width of the broad component must be larger than that of the narrow lines was also applied.

To determine the best-fitted spectral parameters and associated uncertainties, we employed a Monte Carlo bootstrapping approach. We ran 200 simulations and, in each run, the original spectrum was resampled with random normally-distributed noise from the spectral variance and the same fitting model was then applied. The median and median absolute deviation values of the derived spectral parameter distributions were considered as the best fit values and corresponding 1$\sigma$ uncertainties, respectively.

\begin{figure*}
\hbox{
    \includegraphics[width=\linewidth]{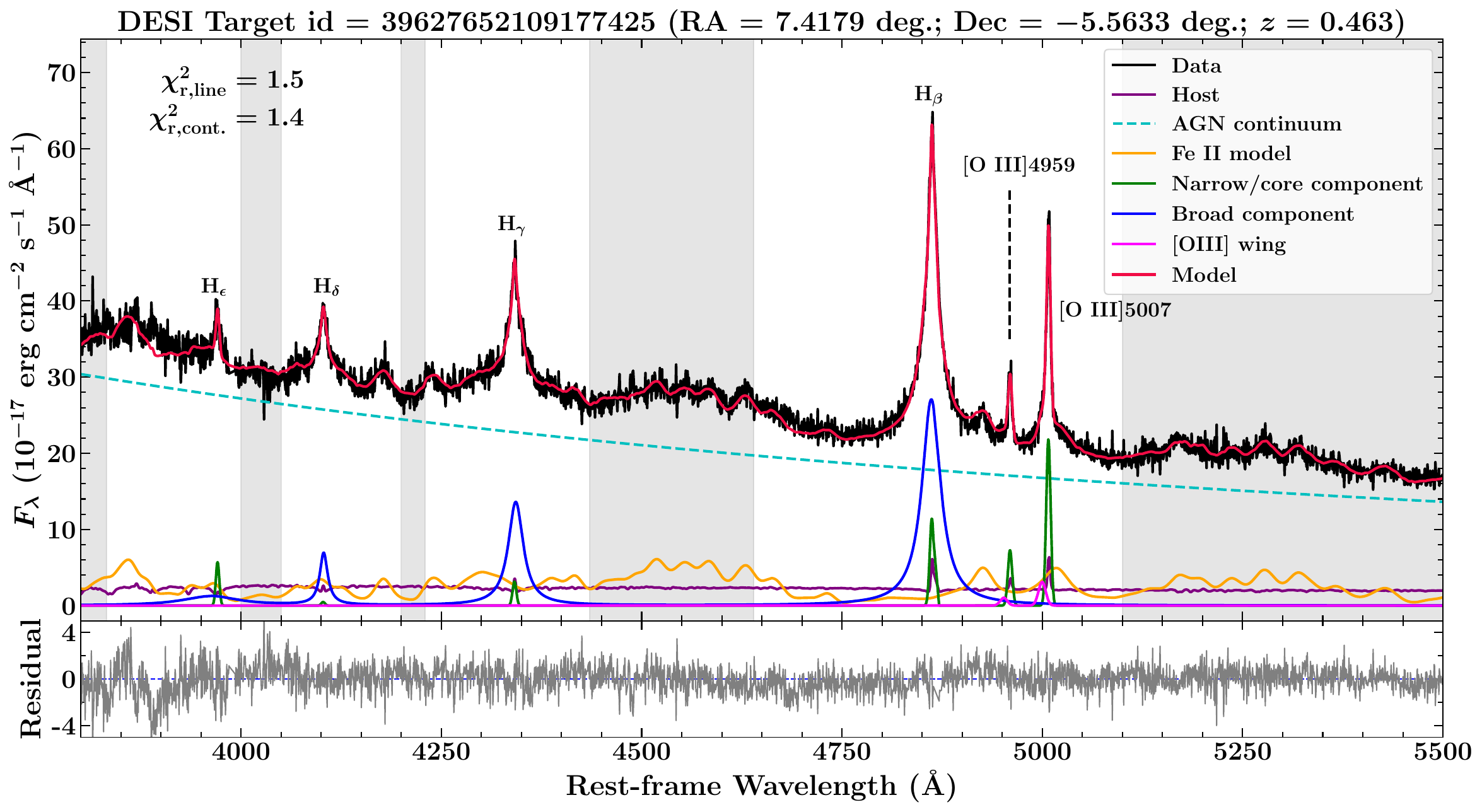}
     }
\caption{This plot shows an example of the spectral decomposition applied to the rest-frame DESI spectrum of an AGN. The model components are labeled, and several emission lines are marked. The vertical shaded regions refer to wavelength regions used for the continuum fitting.}\label{fig:Fig1}
\end{figure*}

\section{Results}\label{sec4}
The spectral fitting was applied on 71918 DESI AGN using the methodology described above. We visually inspected all the fitted spectra to identify sources whose parameters could not be well constrained. The modeling was repeated, when required, for example, without \OIII~wing components to avoid overfitting the \OIII~lines when the outflowing component was not detected. Additionally, we further corrected the redshifts of 170 objects. There were 387 sources with repeated spectra and different DESI target ids. In such cases, we retained the spectrum with higher signal-to-noise ratio. All in all, this exercise led to the exclusion of 16493 optical spectra. Most of them turned out to be either Type 2 AGN or emission line galaxies with no obvious broad \hbeta~component. Other remaining objects have extremely poor quality data precluding us from deriving meaningful parameters. 

Next, we analyzed the spectral parameters of remaining 55425 sources to identify NLSy1 and broad-line Seyfert 1 (BLSy1) galaxies among them. We show an example of the spectral fitting in Figure~\ref{fig:Fig1} and provide the information of the spectral parameters of the sample in the Appendix (Table~\ref{tab:nlsy1_cat}). The parameters are available in the form of NLSy1 and BLSy1 catalogs\footnote{\url{https://www.ucm.es/blazars/seyfert}} at Zenodo (\url{https://doi.org/10.5281/zenodo.20484681}). The reported uncertainties in the spectral parameters, e.g., FWHM, were estimated from the model fitting. Though the uncertainties in the radio flux density values (used for radio-loudness calculation, see, Section~\ref{sec5}) are unavailable, it is typically 3-5\% of the measurements \citep[see, e.g.,][]{2021ApJS..255...30G}. Furthermore, a parameter value of $-$999 refers to the null flag, i.e., the corresponding parameter could not be estimated.

\begin{figure}
\hbox{
    \includegraphics[width=\linewidth]{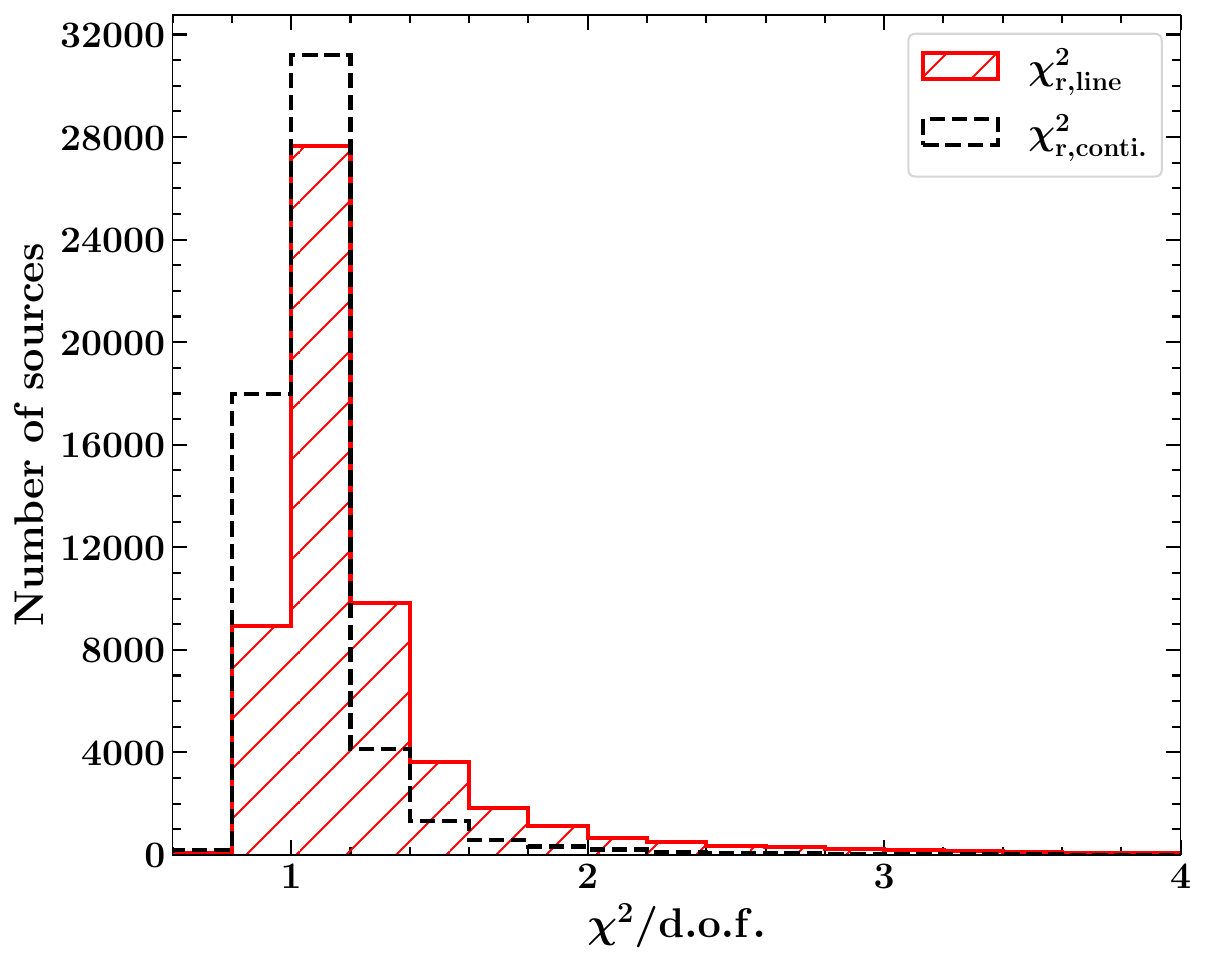}
     }
\caption{The histograms of the reduced-$\chi^2$ estimated from the line ($\chi^2_{\rm r,line}$, red-hatched) and continuum fitting ($\chi^2_{\rm r,conti.}$, black dashed).}\label{fig:chi}
\end{figure}

We estimated the goodness of fit by computing the reduced-$\chi^2$ ($\chi^2_{\rm r}=\chi^2/{\rm dof}$) for the continuum and line fittings and show the corresponding histograms in Figure~\ref{fig:chi}. We found that the spectral fitting pipeline employed in this work reproduces the data reasonably well with over 95\% and 99\% of the fitted sources have $\chi^2_{\rm r}<2$ for the line and continuum fittings, respectively. The visual inspection of some of the objects with large $\chi^2_{\rm r}$ suggested their optical spectra to be of high signal-to-noise ratio. In such objects, the fitting pipeline could not model the data very well. The presence of artifacts in the data, e.g., likely due to imperfect cosmic-ray removal during the data processing, can also affect the $\chi^2_{\rm r}$. However, we visually examined all spectra and confirm that the fitting results are reliable.

For the whole sample, we also calculated bolometric luminosity ($L_{\rm bol}$), $B$-band absolute magnitude ($M_{\rm B}$), single epoch virial black hole mass ($M_{\rm BH}$), and Eddington ratio ($R_{\rm Edd}$). 

We considered the $g$- and $r$-band fluxes, in nanomaggies, estimated by DESI legacy surveys, and converted them to AB-magnitudes using the following formula:

\begin{equation}
    m = 22.5 - 2.5\log_{10}(f),
\end{equation}
where $f$ is the flux in nanomaggy. For 1510 sources, we could not find $g$- and/or $r$-band fluxes in the DESI catalogs. We used the PanSTARRS or SDSS $g$- and $r$-band magnitude for them. For one source (target id: 2781202851299403) we could not find the required magnitude information in literature, likely due to it being in the proximity of a bright star, so it was not considered for $M_{\rm B}$ calculation. The computed $g$- and $r$-band magnitudes were then used to derive the Bessel $B$-band magnitude using the following transformation equation \citep[][]{2006A&A...460..339J}:

\begin{equation}
B = g + (0.313 \pm 0.003) * (g - r) + (0.219 \pm 0.002),
\end{equation}
which was then used to estimate $M_{\rm B}$ using the AGN redshift and $k$-correction. Though most of research works done on NLSy1 galaxies do not differentiate them with narrow-line quasars, the knowledge of $M_{\rm B}$ can be useful for a user to select genuine Seyfert galaxies \citep[$M_{\rm B}>-22.25$,][]{2010A&A...518A..10V}. In the rest of the paper, we do not differentiate between galaxies and quasars and uniformly mention them as NLSy1 or BLSy1 galaxies.

The mass of the central supermassive black hole was calculated assuming the broad line region (BLR) clouds to be virialized, the width of broad lines represent the virial velocity, and the continuum luminosity a proxy for the BLR radius \citep[e.g.,][]{2004MNRAS.352.1390M,2006ApJ...641..689V}. The governing equation can be expressed as follows:

\begin{equation}
M_{\rm BH} = \frac{\kappa R_{\rm BLR} V^2_{\rm BLR}}{G}
\end{equation}
where $G$ is the gravitational constant and the terms $R_{\rm BLR}$ and $V_{\rm BLR}$ refer to the BLR radius and velocity of the BLR clouds, respectively. The geometry and kinematics of the lime emitting BLR clouds are taken into account with the parameter $\kappa$ which was taken as $\kappa=3/4$ for a spherical BLR cloud distribution \citep[cf.][]{2017ApJS..229...39R}. The radius of the BLR, i.e., $R_{\rm BLR}$ was derived using the scaling relation between $R_{\rm BLR}$ and 5100 \AA~continuum luminosity ($\lambda L_{\lambda,5100}$). Following \citet[][]{2019ApJ...886...42D}, the relative strength of \FeII~emission was taken into account since it can have strong implications on the continuum measurement in highly accreting AGN such as NLSy1 sources. We used the following equation to estimate $R_{\rm BLR}$:

\begin{equation}
\log\left(\frac{R_{\rm BLR}}{\rm lt-day}\right) = K_1 + K_2\log\left({\lambda L_{\lambda,5100} \over 10^{44}\,{\rm erg\,s^{-1}}}\right) + K_3 R_{\rm 4570},
\end{equation}
where $R_{\rm 4570}$ is the relative \FeII~strength defined as the ratio of the \FeII~flux in the wavelength range 4434 \AA$-$4684 \AA~to the broad \hbeta~component flux. We adopted the coefficients $K_1$, $K_2$, and $K_3$ as 1.65, 0.45, and $-$0.35 \citep[][]{2019ApJ...886...42D}. We note that the estimated uncertainty in $M_{\rm BH}$ does not consider any systematics inherent to the used scaling relations which can be as large as 0.4 dex \citep[e.g.,][]{2013BASI...41...61S}, and simply corresponds to measurement errors.

The \mbh~measurement in NLSy1 galaxies is an active topic of debate. Since the single-epoch virial technique relies on the FWHM of the broad \hbeta~emission line, NLSy1 objects, with lower FWHM values, tend to have smaller masses of the central black holes. The narrowness of the emission lines hints a relatively slower motion of BLR gas which indicates the presence of a low mass black hole. However, the smaller line widths can also be due to projection effects caused by a `flat' BLR geometry and considering NLSy1 galaxies to be viewed preferentially pole-on \citep[e.g.,][]{2004A&A...426..797C,2008MNRAS.386L..15D,2024Univ...10..254D}. Moreover, \citet[][]{2018NatAs...2...63M} showed that the virial factor used to derive \mbh~is not a constant but depends on line width and/or orientation. Interestingly, a few observational evidence do not support the idea of large \mbh~in NLSy1 galaxies. For example, if these objects are primarily oriented with a pole-on view, they should exhibit a larger radio-loud fraction and faster variability compared to BLSy1 galaxies, due to Doppler boosting effects. However, existing observations do not confirm these predictions \citep[e.g.,][]{2017ApJ...842...96R,2025MNRAS.543..121S}. In the sample of NLSy1s with radio sizes $>$100 kpc, the lack of any correlation of \mbh~with radio core dominance, which is often used as a statistical indicator of orientation, also suggest that low \mbh~may not be due to orientation \citep[][]{2025ApJ...995..125U}. The \mbh~measurements done using X-ray excess variance also suggests low mass black holes powering NLSy1s \citep[e.g.,][]{2001MNRAS.324..653L,2012A&A...542A..83P}. Given the contrasting results found in several works, a dedicated study is needed to ascertain the origin of low \mbh~values found in NLSy1 galaxies. In this work, on the other hand, we adopted the method that has been used in previous NLSy1 catalogs for consistency and a meaningful comparison.

From $\lambda L_{\lambda,5100}$, we derived $L_{\rm bol}$ using the bolometric correction factor of 9.26 following \cite{2006ApJS..166..470R}. We note that the bolometric correction depends on the Eddington ratio as well as observed luminosity \citep[][]{2019MNRAS.488.5185N,2020MNRAS.494.5917F}. Therefore, a constant bolometric correction factor should be used with caution. We also calculated the Eddington ratio ($R_{\rm Edd}$) from $L_{\rm bol}$ and $M_{\rm BH}$.

\subsection{The NLSy1 and BLSy1 Catalogs}
Similar to our previous work, we adopted the following two criteria to identify NLSy1 galaxies in the sample of 55425 objects:

\begin{enumerate}
    \item Considering uncertainties, the FWHM of the broad \hbeta~component is smaller than 2000 \km, i.e., $FWHM-\Delta FWHM \leq2000$ \km.
    \item The ratio of the \OIIIb~and total \hbeta~line fluxes is $<$3, while taking into account the uncertainties in the measured \hbeta~and \OIIIb~flux values.
\end{enumerate}

The first condition of the FWHM threshold of 2000 \km~seems arbitrary, since no sharp differences in the physical properties of AGN have been found around this value. Several works have proposed that a threshold of 4000 \km~is probably more physically motivated as this separates the population A and B sources on the fundamental plane \citep[][]{2000ARA&A..38..521S,2002ApJ...566L..71S,2018FrASS...5....6M}. Therefore, NLSy1 objects can be considered as a sub-sample of Population A sources. However, since the 2000 \km~FWHM threshold has been consistently used to identify and study these sources, we have also chosen the same threshold for a meaningful comparison with earlier works. Furthermore, the use of the FWHM uncertainty ensures that the borderline sources are properly considered.

For the second condition, we considered the fluxes of those narrow \hbeta~and \OIIIb~lines that had non-zero flux uncertainties and signal-to-noise ratio $>$1. Otherwise, we assumed their flux values to be zero since the line detection in such cases is marginal.

The application of above-mentioned two criteria resulted in the identification of 18749 NLSy1 sources present in DESI-DR1. Among them 12968 objects, i.e., $\sim$69\% of the whole sample, are genuine Seyfert galaxies with $M_{\rm B}>-22.25$ \citep[][]{2010A&A...518A..10V}. We found 36 objects that qualified the first criterion but had \OIIIb~to \hbeta~lines flux ratio $>$3. 

We note that several works have reported Seyfert classification schemes based on the \OIIIb~and total \hbeta~line flux ratios. For example, \citet[][]{1992MNRAS.257..677W} adopted the flux ratio upper limit of 0.2, 0.5, and 3 for a Seyfert galaxy of Type 1, 1.2, and 1.5, respectively. Moreover, \citet[][]{1992ApJS...79...49W} proposed a slightly relaxed upper limit of 0.3, 1, and 4, for Seyfert galaxies of Type 1, 1.2, and 1.5, respectively. On the other hand, as per the original classification by \citet[][]{1985ApJ...297..166O}, an NLSy1 has an \OIIIb~and \hbeta~line flux ratio $<$ 3, which is the lower limit for Type 2 Seyfert galaxies \citep[][]{1981ApJ...250...55S}. \citet[][]{1985ApJ...297..166O} briefly noted that sources with the line flux ratio of $\sim$2-3 are more similar to Type 1.5 Seyfert galaxies. Given that all NLSy1 catalogs consistently use the \OIIIb~and \hbeta~line flux ratio threshold of 3, we also adopted the same criterion. However, this implies that a few Type 1.2 or 1.5 Seyfert galaxies could be present in our sample \citep[][]{1992MNRAS.257..677W,1992ApJS...79...49W}. For example, 676 sources, i.e, $\sim$3.6\% of the whole NLSy1 sample, have the \OIIIb~and \hbeta~line flux ratio smaller than 3 but larger than unity. However, none of them have this flux ratio larger than two.

The 36640 BLSy1 sources were selected from the parent sample after excluding 18749 NLSy1 galaxies and 36 narrow line sources that failed to pass the second selection criterion as mentioned above. Among them, 16582 sources are genuine broad line Seyfert 1 AGN with $M_{\rm B}>-22.25$. The remaining 20058 objects can be considered broad emission line quasars ($M_{\rm B}<-22.25$).

\begin{figure*}
\gridline{
    \fig{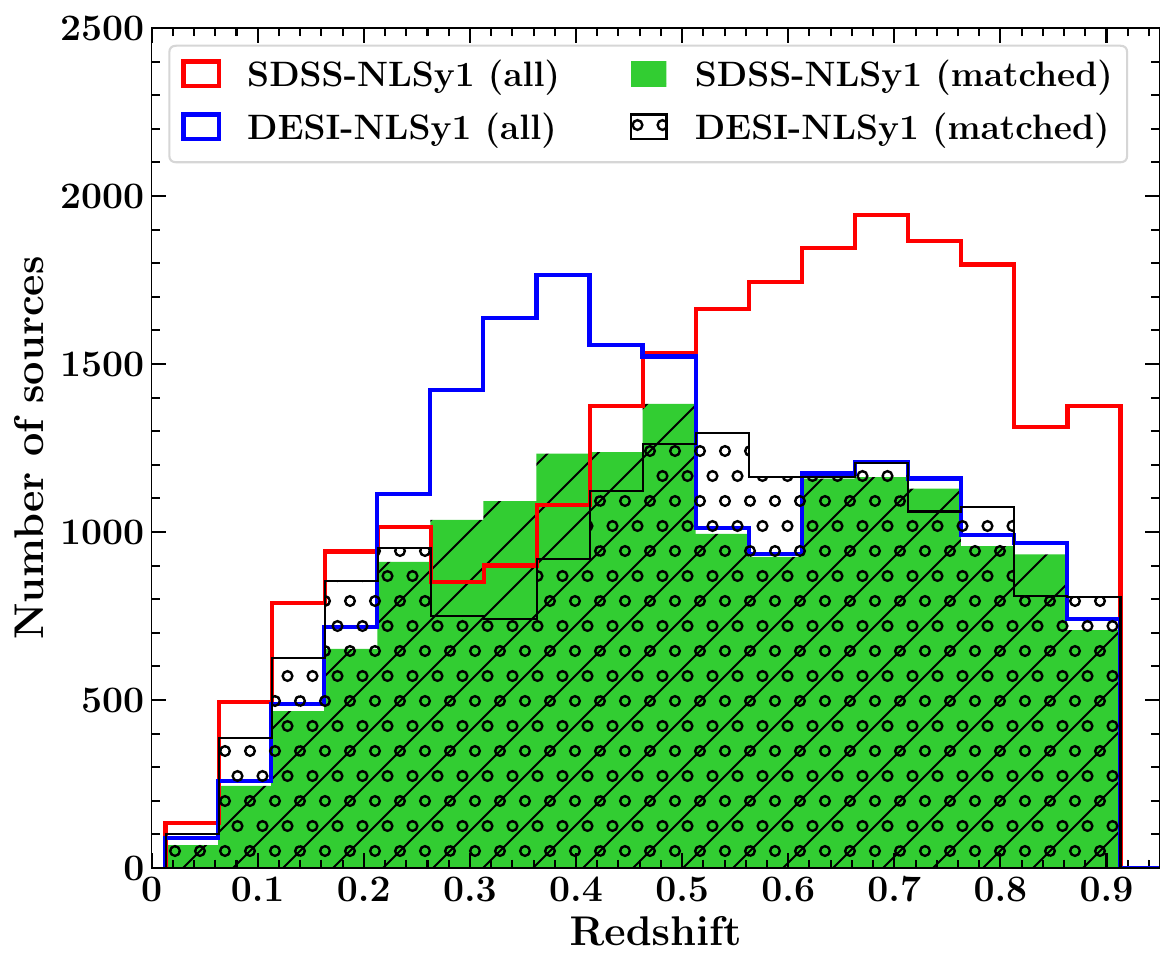}{0.28\textwidth}{(a)}
    \fig{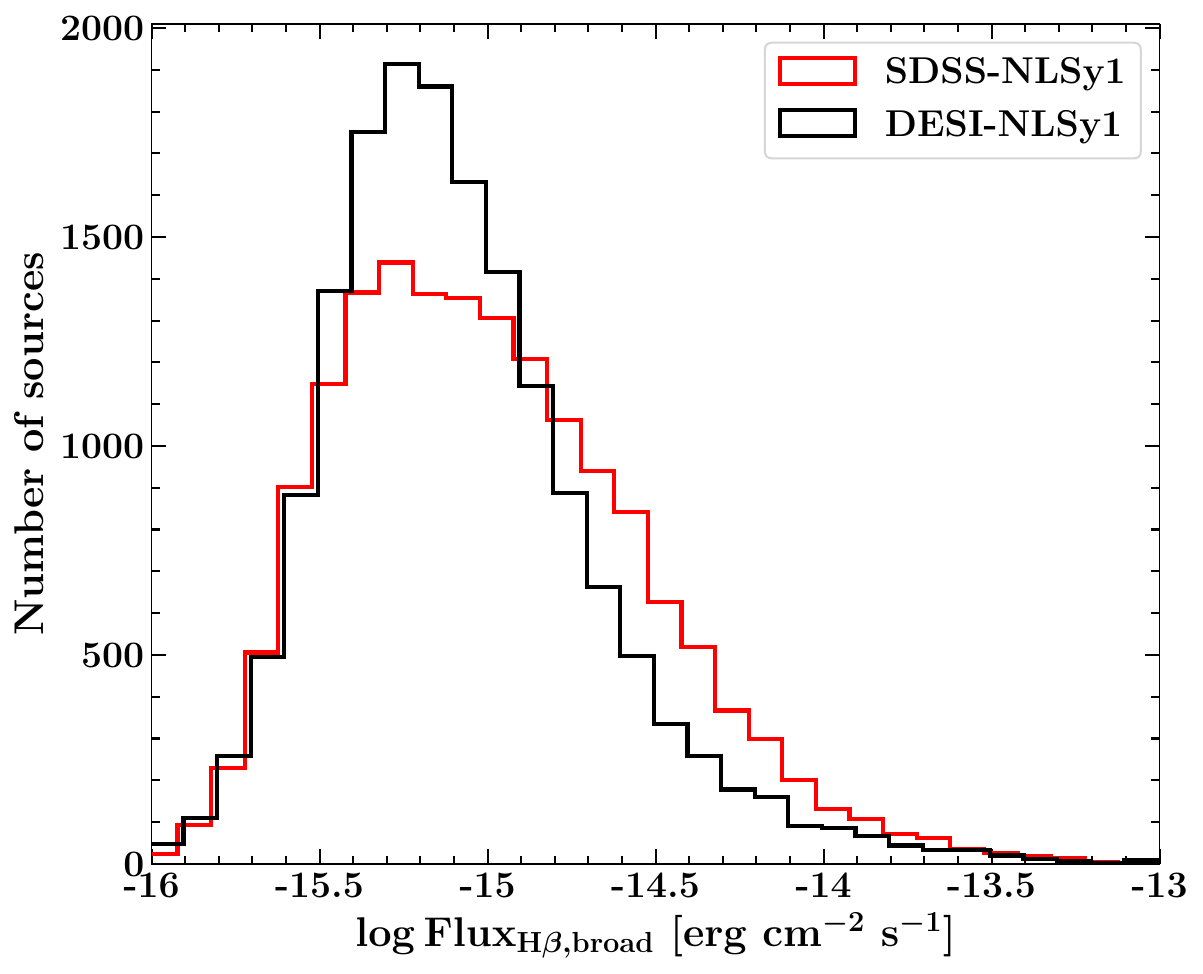}{0.28\textwidth}{(b)}
    \fig{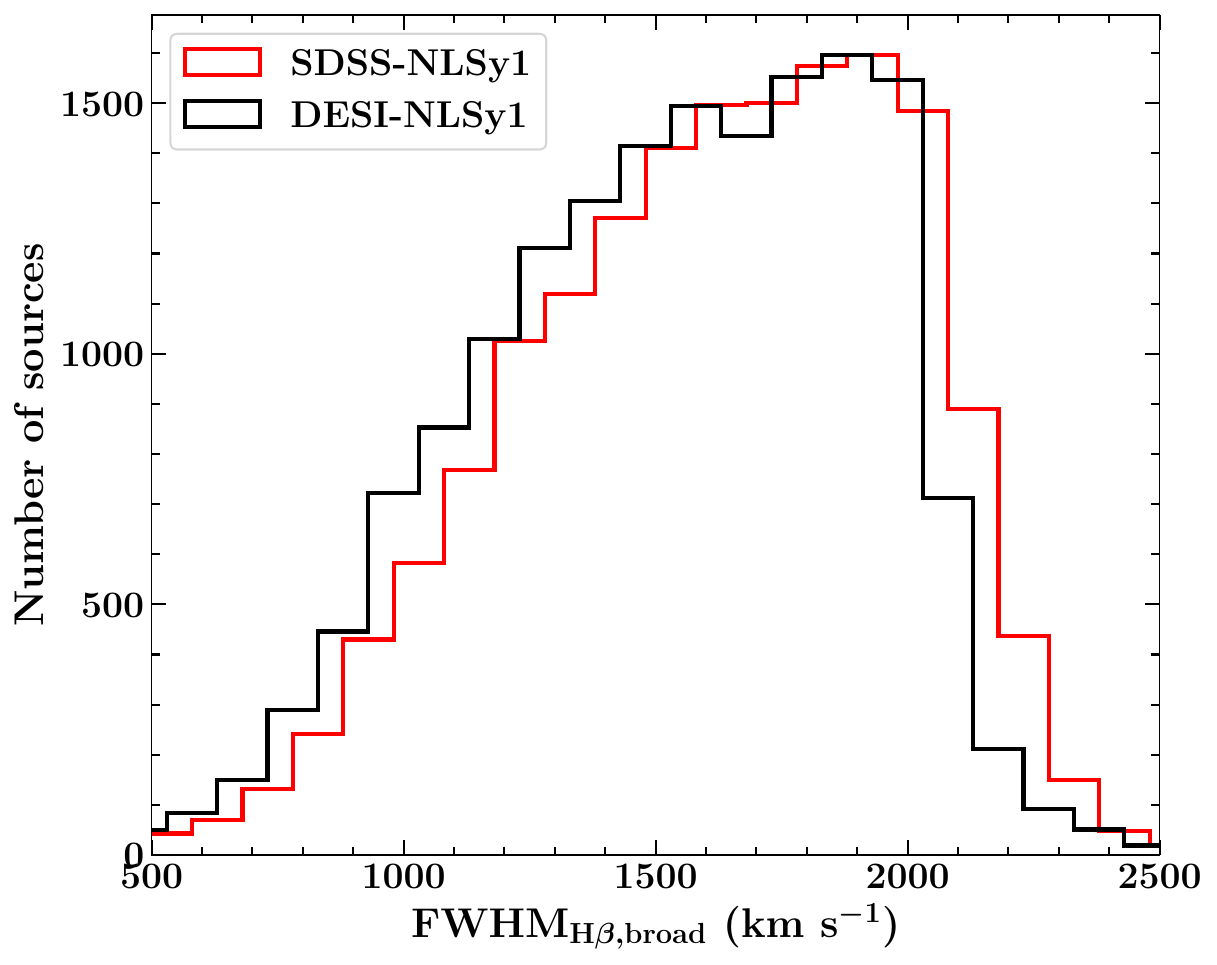}{0.28\textwidth}{(c)}
    }
\gridline{
    \fig{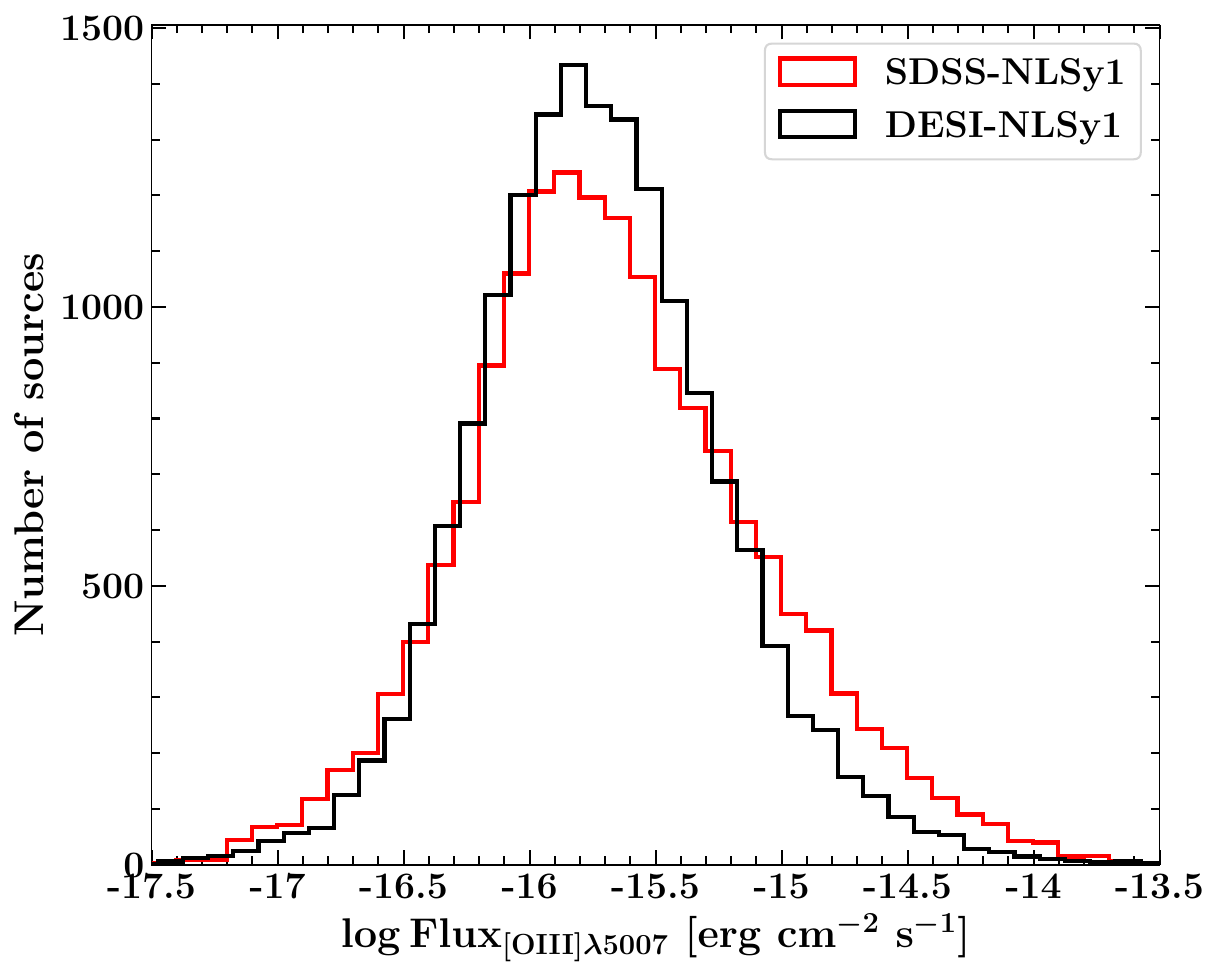}{0.28\textwidth}{(d)}
    \fig{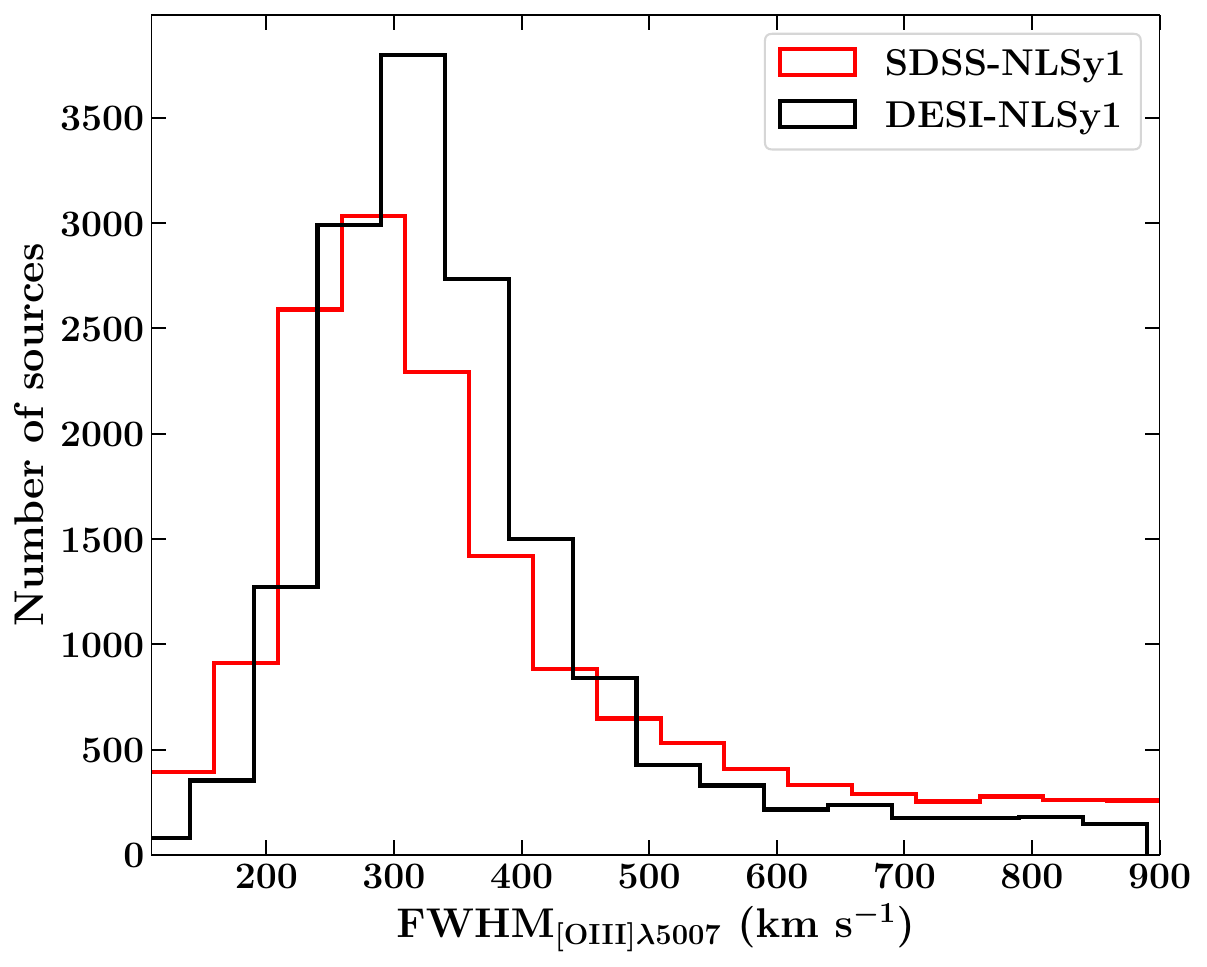}{0.28\textwidth}{(e)}
    \fig{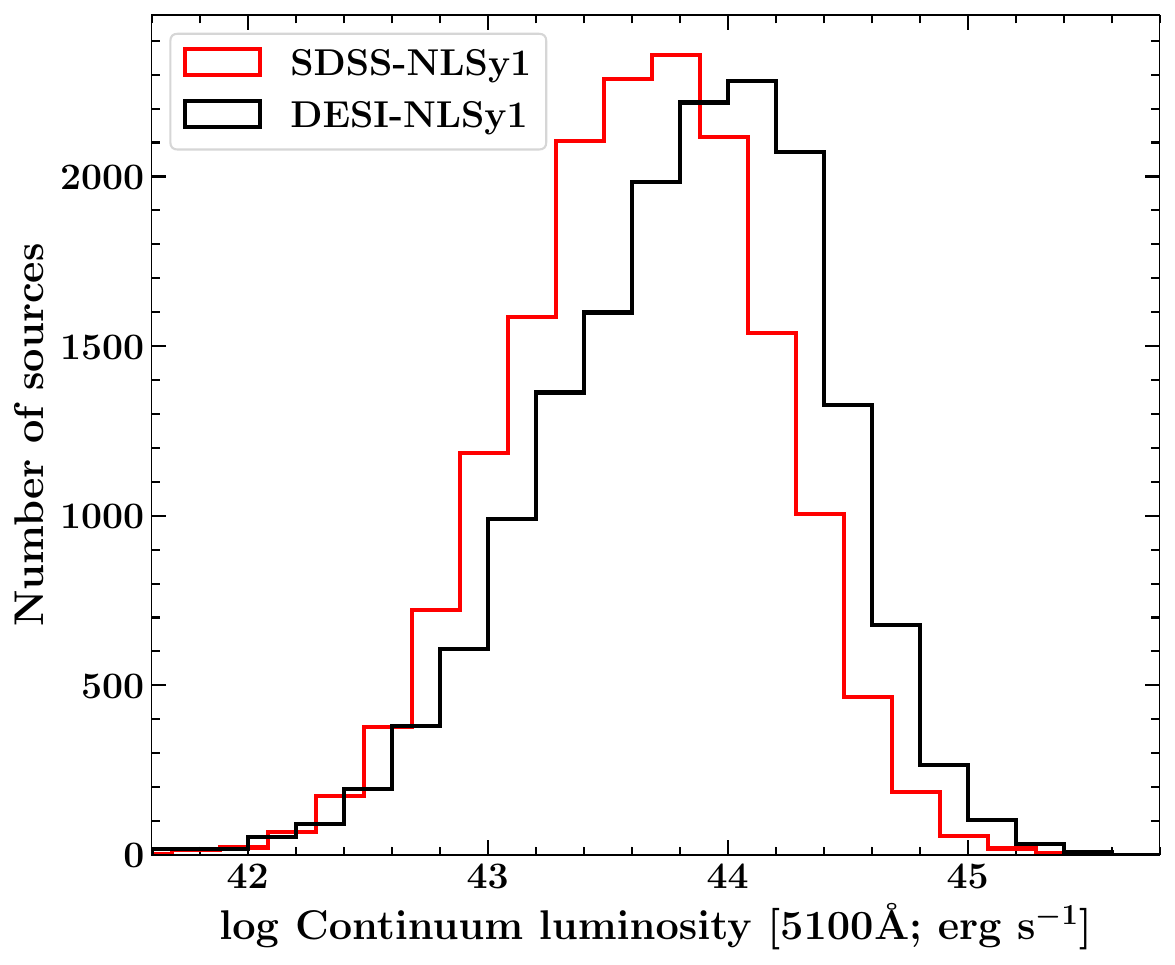}{0.28\textwidth}{(f)}
    }
\gridline{
    \fig{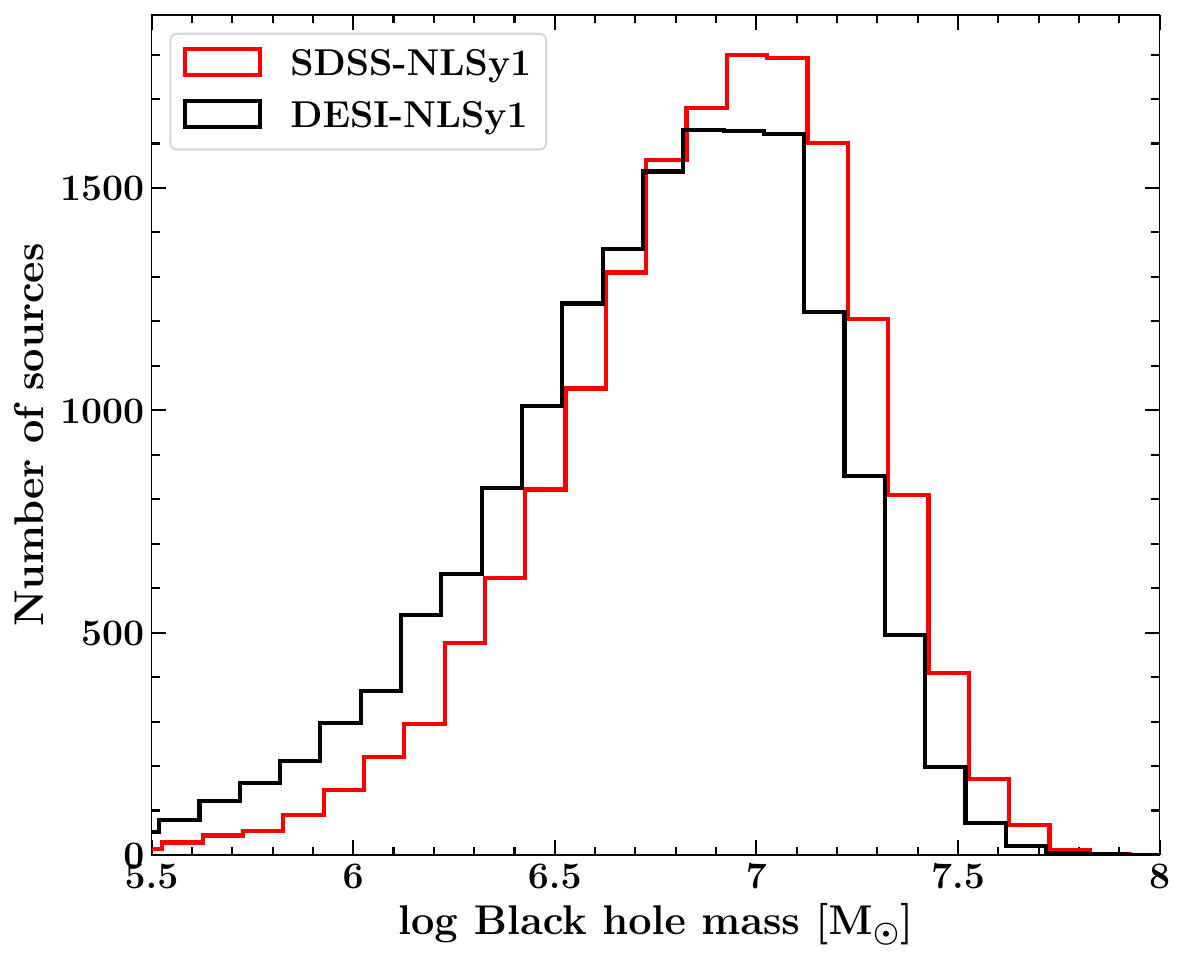}{0.28\textwidth}{(g)}
    \fig{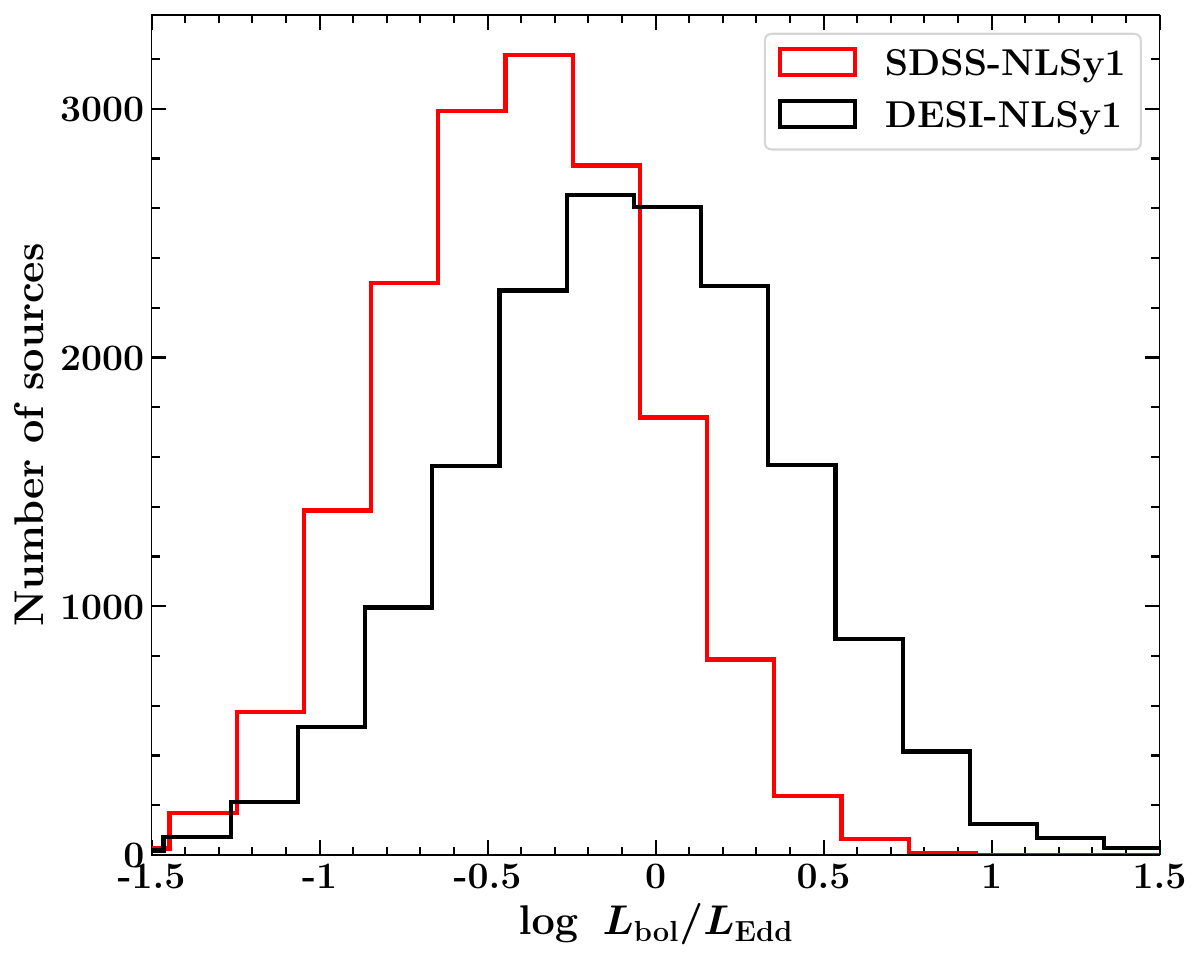}{0.28\textwidth}{(h)}
    \fig{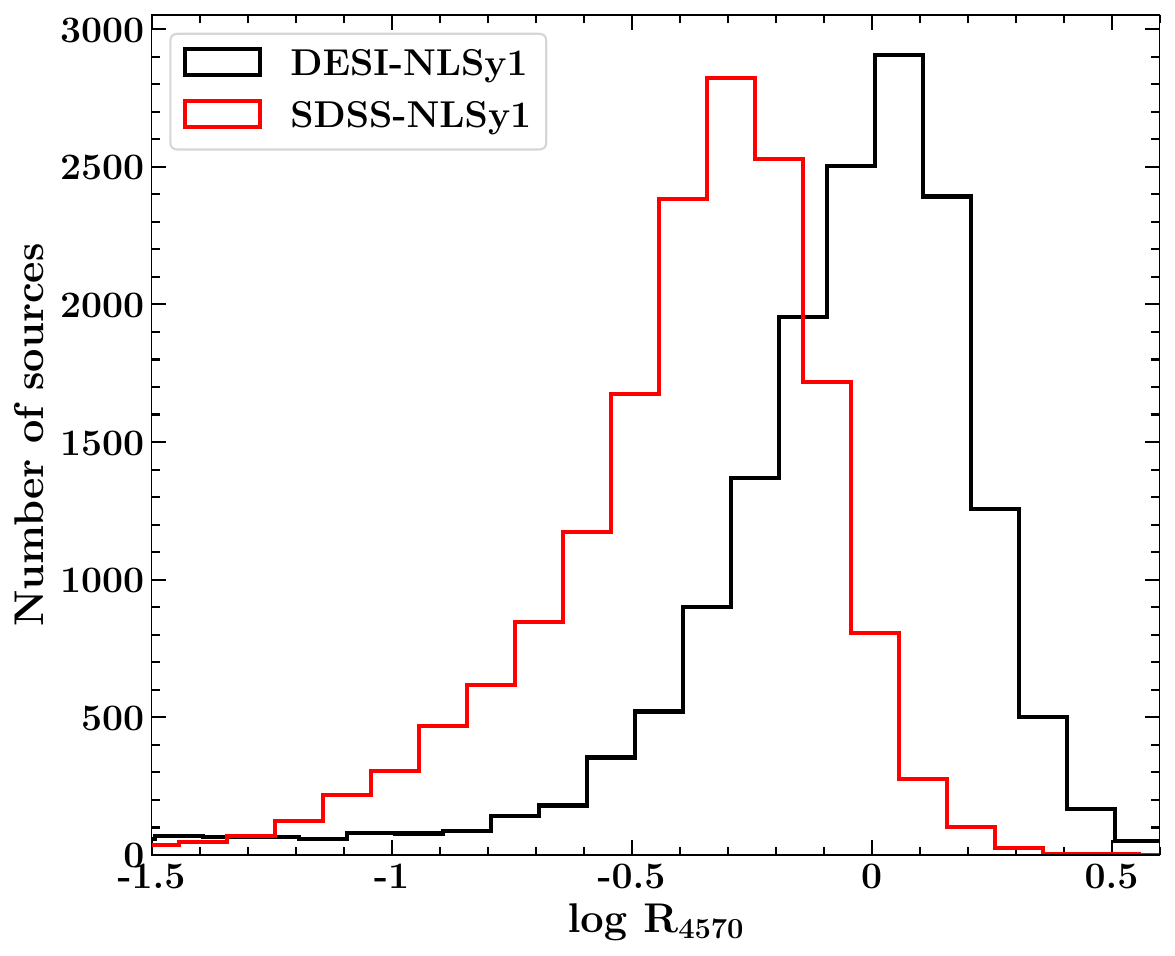}{0.28\textwidth}{(i)}
    }
\caption{These plots show the comparison of several parameters measured/derived from the optical spectroscopic analysis of SDSS-NLSy1 (red) and DESI-NLSy1 (black) galaxies.}\label{fig:comp}
\end{figure*}

\subsection{Comparison with SDSS NLSy1 Galaxies}\label{sec:comp}
We have identified 18749 AGN as NLSy1 galaxies for the first time using the DESI-DR1 data. Since all previous NLSy1 catalogs were made using SDSS datasets, it is intuitive to compare the measured properties of NLSy1s identified in both surveys. For SDSS, we used the NLSy1 catalog based on DR17 \citep[][]{2024MNRAS.527.7055P}. We show the redshift histograms for DESI-NLSy1 and SDSS-NLSy1 sources in panel (a) of Figure~\ref{fig:comp}. The SDSS sample appears to have more high-redshift NLSy1s (median $z=0.58\pm0.17$) compared to those observed with DESI (median $z=0.47\pm0.17$). For a meaningful comparison, we further made a matched selection by choosing SDSS-DESI NLSy1 pairs with similar redshifts ($\Delta z<0.05$) and absolute magnitudes ($\Delta M_{\rm B}<0.1$), finding 16292 pairs. Their redshift distributions are highlighted in panel (a) of Figure~\ref{fig:comp}. Below, we briefly describe the results of the comparison analysis.

{\it Broad \hbeta~Flux}: The comparison of the broad \hbeta~component flux is shown in panel (b) of Figure~\ref{fig:comp}. The overall distributions are similar for DESI-NLSy1 and SDSS-NLSy1 populations with median of the logarithmic flux values $-15.1\pm0.2$ and $-15.0\pm0.3$ (in \ergflux), respectively.

{\it Broad \hbeta~FWHM}: We show the histograms of the FWHM values of the broad \hbeta~component in panel (c) of Figure~\ref{fig:comp}. Similar to the flux distribution, the FWHM also encompasses a similar range for both samples. The median FWHM for the DESI-NLSy1 and SDSS-NLSy1s are $1570\pm285$ \km~and $1650\pm280$ \km, respectively.

{\it \OIIIb~Flux and FWHM}: The distributions of \OIIIb~flux and FWHM are plotted in panels (d) and (e) of Figure~\ref{fig:comp}, respectively. The median flux (log scale, in \ergflux) and FWHM (in \km) values are $-15.7\pm0.3$ and $334\pm62$, and $-15.7\pm0.4$ and $321\pm86$ for DESI-NLSy1 and SDSS-NLSy1 populations, respectively, indicating a similarity in parameters.

{\it Continuum luminosity at 5100 \AA}: The comparison of $\lambda L_{\lambda,5100}$~is shown in panel (f) of Figure~\ref{fig:comp}. The distributions for DESI and SDSS NLSy1s have the median values $43.88\pm0.39$ and $43.65\pm0.36$ (log scale, in \lum), respectively, indicating the former to be slightly more luminous. Moreover, the $L_{\rm bol}$ distribution exhibits a similar trend for both populations, since it was derived from $\lambda L_{\lambda,5100}$.

{\it $M_{\rm BH}$}: The $M_{\rm BH}$ distributions for both source populations are plotted in panel (g) of Figure~\ref{fig:comp}. DESI-NLSy1s tend to have lower black hole masses (median =$6.79\pm0.27$, on log scale, in \msun) compared to the SDSS-NLSy1 objects (median =$6.91\pm0.24$, on log scale, in \msun). However, the overall distributions are similar considering the spread.

{\it Eddington Ratio}: We show the histograms of $R_{\rm Edd}$ ($=L_{\rm bol}/L_{\rm Edd}$) in panel (h) of Figure~\ref{fig:comp}. For DESI-NLSy1 sources, the obtained results reveal a slightly higher $R_{\rm Edd}$ (median = $-0.07\pm0.32$, on log scale), thus higher accretion rate, compared to SDSS-NLSy1 population (median = $-0.40\pm0.27$, on log scale), though the spread is large. 

{\it \FeII~Strength}: We plot the histograms of the \FeII~strength ($R_{\rm 4570}$) in panel (i) of Figure~\ref{fig:comp}. Notably, the DESI-NLSy1 galaxies tend to have considerably larger $R_{\rm 4570}$ compared to SDSS-NLSy1 sources. The median values for both populations are $-0.02\pm0.16$ and $-0.34\pm0.17$, respectively. The observed behavior can be understood within the Eigenvector-1 framework introduced by \citet[][]{1992ApJS...80..109B}, in which NLSy1 galaxies occupy the extreme end characterized by strong optical \FeII~emission and high $R_{\rm Edd}$. The parameter $R_{\rm 4570}$ correlates primarily with $R_{\rm Edd}$ which appeared to be larger for DESI-NLSy1s, and thus explains a higher $R_{\rm 4570}$. Photoionization simulations done with CLOUDY demonstrate that the optical \FeII~strength increases with metallicity in dense BLR clouds. Supersolar metallicity enhances the iron abundance and alters the thermal and ionization structure of the gas, resulting in stronger \FeII~emission relative to H$\beta$ \citep[e.g.,][]{2019ApJ...882...79P}.

\begin{figure}
\hbox{
    \includegraphics[scale=0.4]{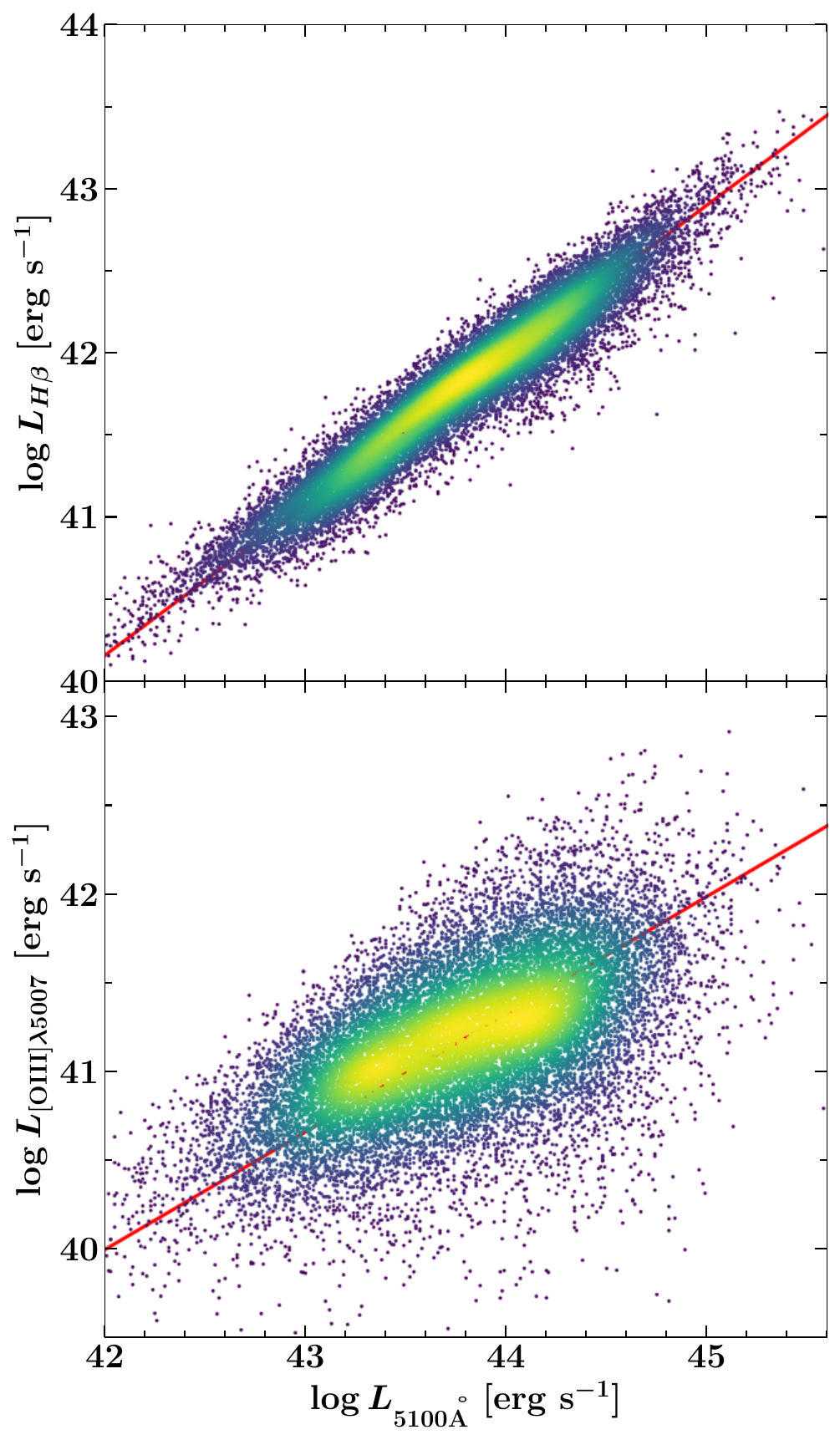}
     }
\caption{The variations of the luminosity of the \hbeta~and \OIIIb~emission lines and 5100\AA~continuum luminosity are shown in these plots. The color scheme is based on the number density of sources with lighter color representing larger number of sources. The red line refers to the best-fitted correlation.}\label{fig:lumin_corr}
\end{figure}

\begin{figure*}
    \hbox{\hspace{1.5cm}
    \includegraphics[scale=0.35]{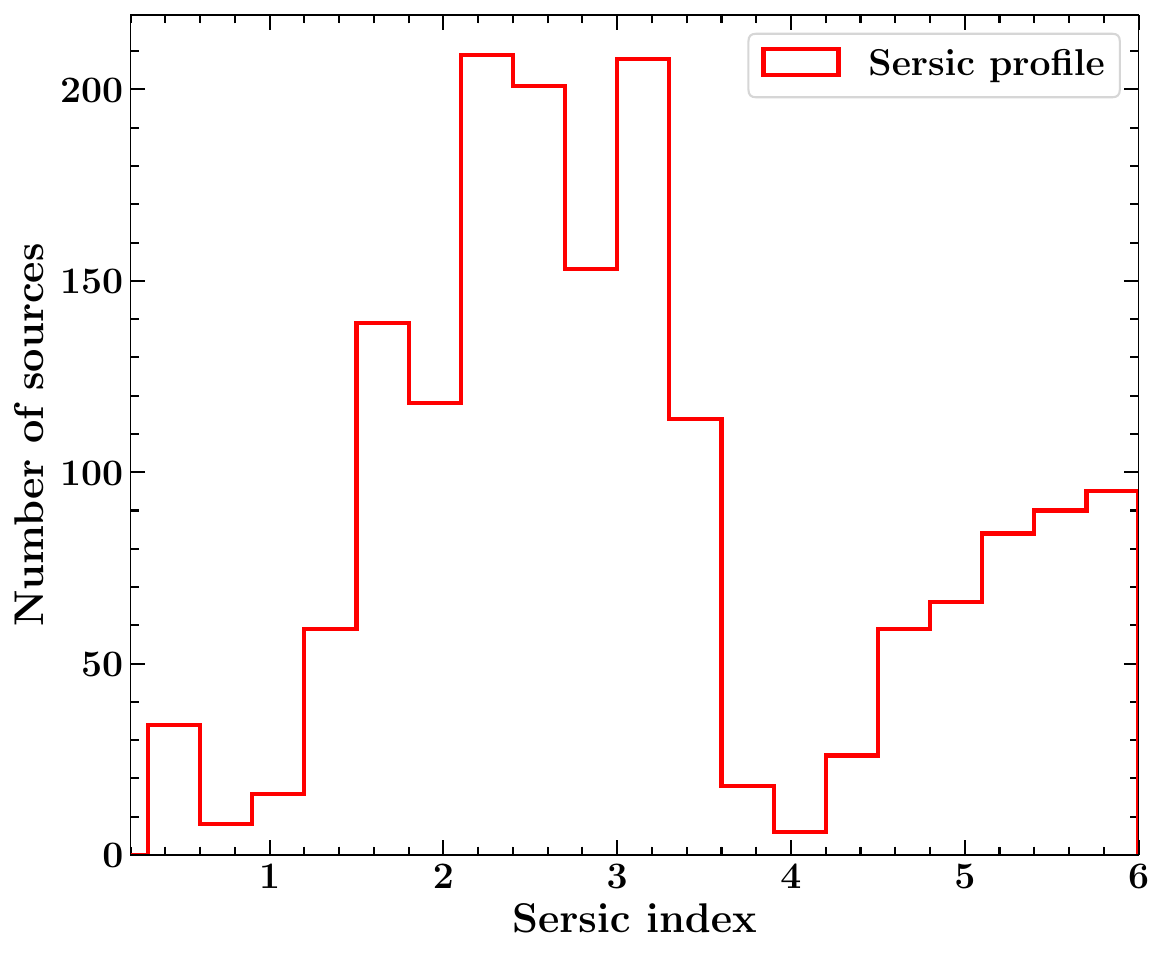}
    \includegraphics[scale=0.35]{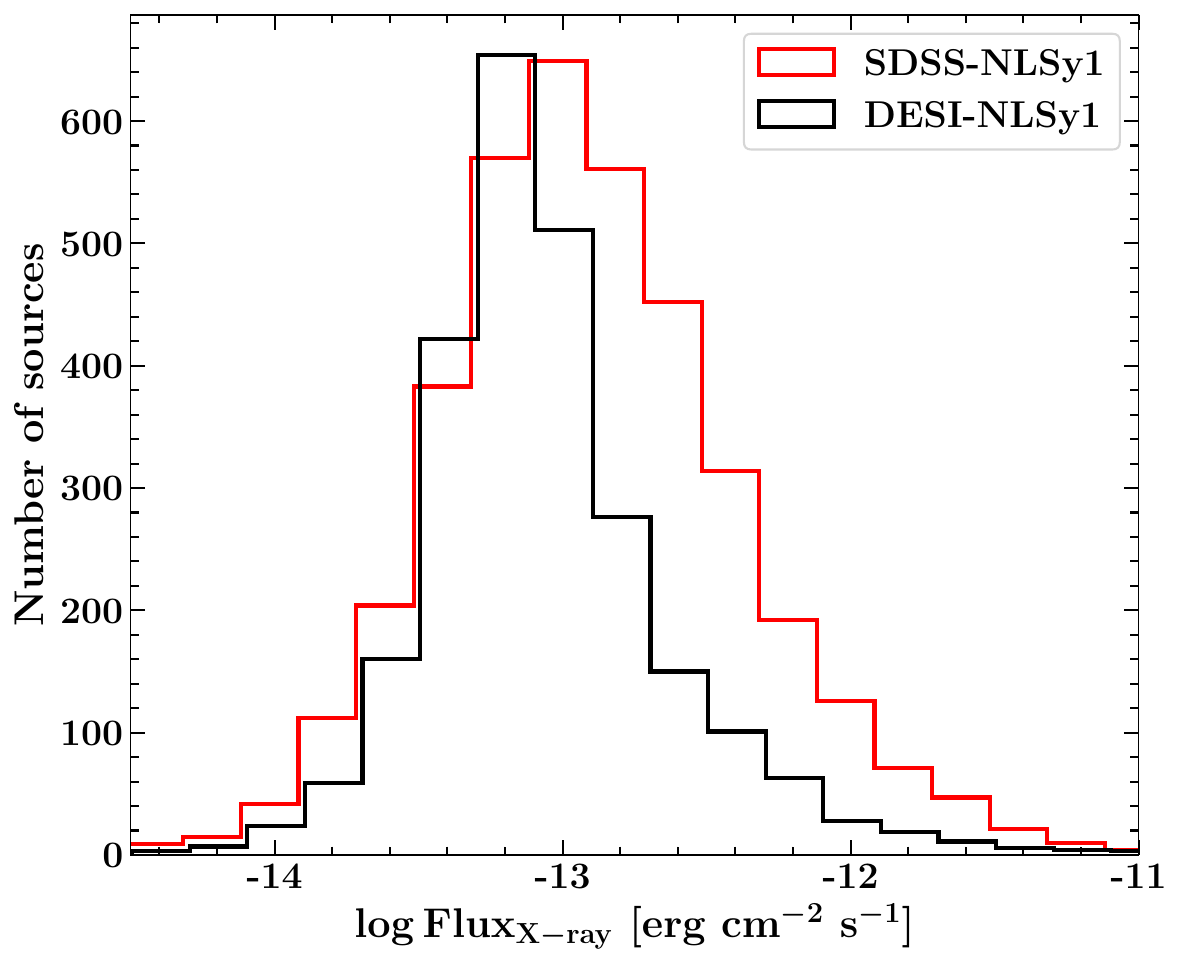}
     }
\caption{Left: The histogram of the S\'{e}rsic index taken from the DESR-DR1 redshift catalog. Right: The distributions of the observed X-ray fluxes for NLSy1 galaxies identified in the DESI (black) and SDSS (red) surveys.}\label{fig:sersic}
\end{figure*}
\section{Multiwavelength Properties}\label{sec5}
We cross-matched the new DESI-NLSy1 catalog with several multi-frequency databases and report our findings below.

\subsection{Radio}
Previous works have reported the detection of only $\sim$5\% NLSy1 to be radio emitters \citep[][]{2006AJ....132..531K,2018MNRAS.480.1796S}. In the SDSS-DR17 NLSy1 catalog, only $\sim$3\% objects were detected in the Faint Images  of the Radio Sky at Twenty-Centimeters survey \citep[FIRST;][]{1997ApJ...475..479W}. In this work, we have used the radio source catalog created by utilizing Very Large Array Sky Survey, an ongoing survey at 2$-$4 GHz covering the whole sky north of $-$40$^{\circ}$ declination \citep[][]{2020PASP..132c5001L,2021ApJS..255...30G}. Since the sky coverage of VLASS is considerably larger than FIRST, this may enable us to identify more radio-emitting NLSy1 sources. We cross-matched the DESI-NLSy1 catalog with the VLASS source catalog made by using the first two epochs of observations and adopting a search radius of 5 arcseconds. We found 285 matches with non-zero integrated flux density values ranging from 0.7 mJy to 503 mJy. Thus, the fraction of radio-detected DESI-NLSy1 is 1.5\%. For an equal comparison, we also cross-matched SDSS-DR17 NLSy1 catalog with the VLASS and found 683 sources, i.e., $\sim$3\% of the whole sample. The lower radio detection fraction of DESI-NLSy1 galaxies could be due to these being relatively low-luminosity galaxies as radio AGN activity has been shown to have a strong dependence on galaxy mass \citep[cf.][]{2019A&A...622A..17S}, which in turn depends on the absolute magnitude \citep[e.g.][]{2005ApJ...635..260S}. 

We also estimated the radio-loudness parameter ($R$) which is defined as the ratio of the rest-frame 5 GHz flux density and 4400 \AA~\citep[][]{1989AJ.....98.1195K}. We extrapolated the 3 GHz flux density measured by VLASS to 5 GHz assuming the spectral index $\alpha=-0.5$ ($F_{\nu}\propto\nu^{\alpha}$). A total of 230 NLSy1 objects, i.e., $\sim$81\% of all radio detected ones, were found to be radio-loud ($R>10$) which is larger than 63\% found for SDSS-NLSy1 sample \citep[][]{2024MNRAS.527.7055P}. It is relevant to note that while studying the radio properties of the sample of NLSy1s compiled by \cite{2017ApJS..229...39R}, \cite{2018MNRAS.480.1796S} found $\sim$81.7\% to be radio-loud, similar to our sample of DESI-NLSy1s. The sample of SDSS-NLSy1s extends to higher redshifts, and it would be useful to make deeper observations to check whether radio-loud objects may have been missed in this sample.

\subsection{Infrared}
We cross-matched the DESI-DR1 NLSy1 catalog with the allWISE source catalog \citep[][]{2019ipac.data...I1W} with a search radius of 5 arcseconds. In $W1$-filter, we found 17801 detections at $\geq3\sigma$ confidence level, i.e., $\sim$95\% of the whole sample which is similar to $\sim$96\% found for the SDSS-NLSy1 catalog. On the other hand, 17725 ($\sim$95\%) and 12909 ($\sim$69\%) NLSy1s were identified in $W2$ and $W3$ bands, respectively. This implied a lower $W3$-band detection compared to $\sim$80\% found for SDSS-NLSy1 sample \citep[][]{2024MNRAS.527.7055P}.

\subsection{Optical}
Several previous works reported a strong correlation between emission line luminosities and the continuum luminosity for AGN, including NLSy1 galaxies \citep[][]{2005ApJ...630..122G,2006ApJS..166..128Z,2017ApJS..229...39R}. We examined these correlations for the DESI-NLSy1 sample by applying a simple linear least-square fit to the data. We obtained the following empirical relations (Figure~\ref{fig:lumin_corr}):

\begin{equation}
\log(L_{\mathrm{H\beta}})=(1.769 \pm 0.085)+ (0.914 \pm 0.002) \log L_{5100}.
\end{equation} 
\begin{equation}
\log(L_{\mathrm{[O {\small III}]\lambda5007}})= (12.161 \pm 0.218) + (0.663 \pm 0.005) \log L_{5100}.
\end{equation} 
These results are similar to those reported in literature \citep[e.g.,][]{2024MNRAS.527.7055P}.

The DESI-DR1 redshift catalog provides a measure of the galaxy morphology by fitting several models listed as follows\footnote{\url{https://www.legacysurvey.org/dr9/catalogs/}}: point source (type=``PSF''), round exponential galaxy model (``REX''), de Vaucouleurs model (``DEV''), exponential model (``EXP''), and a S\'{e}rsic model (``SER''). Among them, the commonly adopted S\'{e}rsic model can give clues about the galaxy light profile based on the parameter S\'{e}rsic index, also provided in the DESI-DR1 redshift catalog. Admittedly, the DESI legacy survey data modeling was done following an automated process; however, the estimated S\'{e}rsic index can give us clues about the host galaxy morphology of the overall NLSy1 population. Among 18749 DESI-DR1 NLSy1 galaxies, the S\'{e}rsic model is the best-fitted profile for 3228 sources. Excluding objects with S\'{e}rsic index value of 6, i.e., the upper limit, we had 1699 NLSy1 galaxies. We show the distribution of their S\'{e}rsic index in the left panel of Figure~\ref{fig:sersic}. As can be seen, most of the sources have S\'{e}rsic index smaller than 4 which implies their host galaxies to be late-type, disk-dominated (for S\'{e}rsic index $\lesssim$2) or intermediate types (S0/Sb) with both disk and bulges. A large S\'{e}rsic index ($>$4), on the other hand, indicates an early-type, elliptical host. Since finding a NLSy1 nucleus in an elliptical host is uncommon, we examined the DESI legacy survey images of some of the sources with large S\'{e}rsic index. In several cases (e.g., target id: 39633342278142022), we found spiral arms and complex structures, thereby suggesting that a large S\'{e}rsic index may be due to poor fitting. All in all, these results support the notion that NLSy1 sources are generally hosted in late-type, non-elliptical galaxies \citep[e.g.,][]{2003AJ....126.1690C,2016MNRAS.460.3202O}.

\subsection{X-ray}
To determine the X-ray detected NLSy1 galaxies, using a search radius of 5 arcseconds, we scanned the \chandra~source catalog \citep[CSC v2.1, 0.5$-$7 keV;][]{2024ApJS..274...22E}, the first X-ray catalogs and data release of the western Galactic hemisphere by the SRG/eROSITA all-sky survey \citep[eRASS-DR1, 0.2$-$2.3 keV;][]{2024A&A...682A..34M}, \xmm~serendipitous source catalog \citep[4XMM-DR14, 0.2$-$12 keV;][]{2020A&A...641A.136W}, and live \swift-X-ray Telescope Point Source Catalog \citep[LSXPS, 0.3$-$10 keV;][]{2023MNRAS.518..174E}, in sequential order. This exercise led to identification of 264, 1388, 443, and 417 X-ray emitting NLSy1 galaxies, respectively. We show the X-ray flux measured by the mentioned surveys, albeit having overlapping yet different energy ranges, in the right panel of Figure~\ref{fig:sersic}. For a comparison, we also plot the X-ray flux histogram for SDSS-DR17 NLSy1 galaxies using the flux values reported by the same catalogs mentioned above. The DESI-NLSy1 objects are slightly fainter (log scale median flux = $-13.12\pm0.22$, in \ergflux) compared to the SDSS-NLSy1s (log scale median flux = $-12.94\pm0.32$, in \ergflux) though the spreads of the distributions are large.

\subsection{Gamma-ray}
The presence of closely aligned relativistic jets in some of the radio-loudest NLSy1 galaxies was confirmed by their detection with the Fermi Large Area Telescope \citep[cf.][]{2019JApA...40...39P}. The \gm-ray detection also provides strong evidence that AGN powered by low black hole masses can also host powerful relativistic jets akin to blazars. To identify \gm-ray emitting NLSy1 galaxies in the DESI sample, we utilized the recently released early 16-year list of Fermi-LAT detected sources \citep[FL16Y;][]{2022ApJS..260...53A,2026arXiv260222148B}. We cross-matched both catalogs using a 5 arcseconds search radius and adopting the counterpart coordinates provided in FL16Y. We found 2 \gm-ray emitting sources: FL16Y J0017.5$-$0512 and FL16Y J1754.8+3443. These objects are associated with radio sources PMN J0017$-$0512 ($z=0.227$) and MG2 J175448+3442 ($z=0.62$), respectively. Such a small number of \gm-ray emitting DESI-NLSy1 galaxies is surprising. This is because the SDSS-NLSy1 sample, which has a slightly larger size of 22656 NLSy1s compared to 18756 DESI-NLSy1s, has 21 \gm-ray detected objects \citep[][]{2024MNRAS.527.7055P}. One possible reason to explain the differences could be the fact that DESI-NLSy1s are intrinsically less powerful AGN compared to SDSS-NLSy1 galaxies. This hypothesis can also explain their lower X-ray fluxes (Figure~\ref{fig:sersic}), fainter $M_{\rm B}$, and lower black hole masses (Figure~\ref{fig:comp}). Deeper multiwavelength investigations of these objects are needed to probe their enigmatic behavior.

\section{Summary}\label{sec6}
We have utilized the recently released optical spectroscopic data from the DESI-DR1 catalog to identify NLSy1 galaxies. We started with the AGN/QSO VAC released by the DESI collaboration, and applied a number of filters to remove contaminating objects. In the wavelength range of 3800 \AA$-$5500 \AA, we subtracted the continuum by modeling the rest-frame optical spectrum with a combination of the nuclear power-law AGN emission, \FeII~template, and prior-informed host galaxy templates generated using PCA analysis. The continuum subtracted spectrum was reproduced with a combination of Lorentzian (for the broad \hbeta~component) and Gaussian (for the narrow components) functions, and spectral parameters were estimated by adopting a Monte Carlo bootstrapping technique. This exercise resulted in a new catalog of 18749 NLSy1 galaxies identified for the first time. Several of their optical spectroscopic parameters or those derived from them are similar to those estimated for previously known NLSy1 objects. However, the detection fractions of DESI-NLSy1 galaxies in radio, X-ray, and \gm-ray surveys were found to be lower than SDSS-NLSy1 objects. These intriguing findings indicate that NLSy1s identified in the DESI survey may belong to the low-luminosity members of this class of AGN. Deeper broadband observations of these enigmatic sources will play a pivotal role in exploring the physics of AGN at the faint end of the source population, thus setting benchmark for their observations with the next-generation of large telescopes.

\acknowledgements
We thank the journal referee for constructive criticism. A.D. is thankful for the support of the Ram{\'o}n y Cajal program from the Spanish MINECO, Proyecto PID2021-126536OA-I00 funded by MCIN / AEI / 10.13039/501100011033, and Proyecto PR44/21‐29915 funded by the Santander Bank and Universidad Complutense de Madrid.

This research used data obtained with the Dark Energy Spectroscopic Instrument (DESI). DESI construction and operations is managed by the Lawrence Berkeley National Laboratory. This material is based upon work supported by the U.S. Department of Energy, Office of Science, Office of High-Energy Physics, under Contract No. DE–AC02–05CH11231, and by the National Energy Research Scientific Computing Center, a DOE Office of Science User Facility under the same contract. Additional support for DESI was provided by the U.S. National Science Foundation (NSF), Division of Astronomical Sciences under Contract No. AST-0950945 to the NSF's National Optical-Infrared Astronomy Research Laboratory; the Science and Technology Facilities Council of the United Kingdom; the Gordon and Betty Moore Foundation; the Heising-Simons Foundation; the French Alternative Energies and Atomic Energy Commission (CEA); the National Council of Humanities, Science and Technology of Mexico (CONAHCYT); the Ministry of Science and Innovation of Spain (MICINN), and by the DESI Member Institutions: www.desi.lbl.gov/collaborating-institutions. The DESI collaboration is honored to be permitted to conduct scientific research on I'oligam Du'ag (Kitt Peak), a mountain with particular significance to the Tohono O'odham Nation. Any opinions, findings, and conclusions or recommendations expressed in this material are those of the author(s) and do not necessarily reflect the views of the U.S. National Science Foundation, the U.S. Department of Energy, or any of the listed funding agencies.
This publication makes use of data products from the Wide-field Infrared Survey Explorer, which is a joint project of the University of California, Los Angeles, and the Jet Propulsion Laboratory/California Institute of Technology, funded by the National Aeronautics and Space Administration.

This research has made use of the NASA/IPAC Extragalactic Database (NED), which is operated by the Jet Propulsion Laboratory, California Institute of Technology, under contract with the National Aeronautics and Space Administration. Part of this work is based on archival data, software or online services provided by the Space Science Data Center (SSDC). This research has made use of NASA's Astrophysics Data System Bibliographic Services.

\appendix
\section{Details of the Catalog Fits File. The null flag is $-$999 for all columns.}\label{sec:app}
We provide the full description of the fits catalog below in Table~\ref{tab:nlsy1_cat}.

\begin{table*}
\caption{The format of the FITS catalog. \label{tab:nlsy1_cat}}
\begin{center}
\begin{tabular}{llll}
\hline
Column  & Format & Units & Description \\
\hline
NAME & STRING &  & J2000 designation \\
TARGET\_ID              & STRING  &  & Spectroscopic Plate number, MJD, and fiber number \\
RA                    & DOUBLE & degree & Right ascension (J2000) \\
DEC                 & DOUBLE & degree & Declination (J2000) \\
Z                      & DOUBLE &      & Spectroscopic redshift \\
M\_B               & DOUBLE  &     & Absolute $B$-band magnitude \\
BR\_H\_BETA\_FWHM & DOUBLE & \km  & FWHM of the broad \hbeta~emission line \\
BR\_H\_BETA\_FWHM\_ERR & DOUBLE & \km  & Uncertainty in the FWHM of the broad \hbeta~emission line \\
BR\_H\_BETA\_FLUX & DOUBLE & [\ergflux]  & Flux of the broad \hbeta~emission line on log scale \\
BR\_H\_BETA\_FLUX\_ERR & DOUBLE & [\ergflux] & Uncertainty in the log-scale flux of the broad \hbeta~emission line \\
BR\_H\_BETA\_LUM & DOUBLE & [\lum]  & Luminosity of the broad \hbeta~emission line on log scale \\
BR\_H\_BETA\_LUM\_ERR & DOUBLE & [\lum] & Uncertainty in the log-scale luminosity of the broad \hbeta~emission line \\
NA\_OIII\_5007\_FWHM & DOUBLE & \km  & FWHM of the narrow \OIIIb~emission line \\
NA\_OIII\_5007\_FWHM\_ERR & DOUBLE & \km  & Uncertainty in the FWHM of the narrow \OIIIb~emission line \\
NA\_OIII\_5007\_FLUX & DOUBLE & [\ergflux]  & Flux of the narrow \OIIIb~emission line on log scale \\
NA\_OIII\_5007\_FLUX\_ERR & DOUBLE & [\ergflux] & Uncertainty in the log-scale flux of the narrow \OIIIb~emission line \\
NA\_OIII\_5007\_LUM & DOUBLE & [\lum]  & Luminosity of the narrow \OIIIb~emission line (\lum) on log scale \\
NA\_OIII\_5007\_LUM\_ERR & DOUBLE & [\lum] & Uncertainty in the log-scale luminosity of the narrow \OIIIb~emission line \\
NA\_H\_BETA\_FLUX & DOUBLE &  [\ergflux] & Flux of the narrow \hbeta~emission line on log scale \\
NA\_H\_BETA\_FLUX\_ERR & DOUBLE & [\ergflux] & Uncertainty in the log-scale flux of the narrow \hbeta~emission line \\
NA\_H\_BETA\_LUM & DOUBLE & [\lum]  & Luminosity of the narrow \hbeta~emission line on log scale \\
NA\_H\_BETA\_LUM\_ERR & DOUBLE & [\lum] & Uncertainty in the log-scale luminosity of the narrow \hbeta~emission line \\
L5100 & DOUBLE & [\lum] & Rest-frame AGN continuum luminosity at 5100 \AA~on log scale \\
L5100\_ERR & DOUBLE & [\lum] & Uncertainty in the log scale continuum luminosity at 5100 \AA \\
L5100\_HOST & DOUBLE & [\lum] & Rest-frame Host continuum luminosity at 5100 \AA~on log scale \\
R4570                         & DOUBLE  &   & \FeII~strength \\
SN\_CONTI                         & DOUBLE  &   & Continuum signal-to-noise ratio\\
M\_BH & DOUBLE & [\msun] & Black hole mass on log scale \\
M\_BH\_ERR & DOUBLE & [\msun] & Uncertainty in the log-scale black hole mass \\
L\_BOL & DOUBLE & [\lum] & Bolometric luminosity on log scale \\
L\_BOL\_ERR & DOUBLE & [\lum] & Uncertainty in the log-scale bolometric luminosity\\
R\_EDD & DOUBLE &  & Eddington ratio on log scale \\
R\_EDD\_ERR & DOUBLE &  & Uncertainty in the log-scale Eddington ratio\\
RL & DOUBLE & & radio-loudness parameter on log scale \\
\hline
\end{tabular}
\end{center}
\end{table*}

\bibliographystyle{aasjournal}
\bibliography{Master}

\begin{thebibliography}{}
\expandafter\ifx\csname natexlab\endcsname\relax\def\natexlab#1{#1}\fi

\bibitem[{{Abdo} {et~al.}(2009){Abdo}, {Ackermann}, {Ajello}, {Baldini},
  {Ballet}, {Barbiellini}, {Bastieri}, {Bechtol}, {Bellazzini}, {Berenji},
  {Bloom}, {Bonamente}, {Borgland}, {Bregeon}, {Brez}, {Brigida}, {Bruel},
  {Burnett}, {Caliandro}, {Cameron}, {Caraveo}, {Casandjian}, {Cecchi}, {{\c
  C}elik}, {Chekhtman}, {Cheung}, {Chiang}, {Ciprini}, {Claus}, {Cohen-Tanugi},
  {Conrad}, {Cutini}, {Dermer}, {de Palma}, {Silva}, {Drell}, {Dubois},
  {Dumora}, {Farnier}, {Favuzzi}, {Fegan}, {Focke}, {Foschini}, {Frailis},
  {Fukazawa}, {Fusco}, {Gargano}, {Gehrels}, {Germani}, {Giebels}, {Giglietto},
  {Giordano}, {Giroletti}, {Glanzman}, {Godfrey}, {Grenier}, {Grove},
  {Guillemot}, {Guiriec}, {Hayashida}, {Hays}, {Horan}, {Hughes},
  {J{\'o}hannesson}, {Johnson}, {Johnson}, {Kadler}, {Kamae}, {Katagiri},
  {Kataoka}, {Kerr}, {Kn{\"o}dlseder}, {Kuss}, {Lande}, {Latronico}, {Longo},
  {Loparco}, {Lott}, {Lovellette}, {Lubrano}, {Makeev}, {Mazziotta},
  {McConville}, {McEnery}, {Meurer}, {Michelson}, {Mitthumsiri}, {Mizuno},
  {Monte}, {Monzani}, {Morselli}, {Moskalenko}, {Murgia}, {Nolan}, {Norris},
  {Nuss}, {Ohsugi}, {Omodei}, {Orlando}, {Ormes}, {Pelassa}, {Pepe}, {Persic},
  {Pesce-Rollins}, {Piron}, {Porter}, {Rain{\`o}}, {Rando}, {Razzano},
  {Rochester}, {Rodriguez}, {Ryde}, {Sadrozinski}, {Sambruna}, {Sander}, {Saz
  Parkinson}, {Scargle}, {Sgr{\`o}}, {Smith}, {Spandre}, {Spinelli},
  {Strickman}, {Suson}, {Tagliaferri}, {Takahashi}, {Takahashi}, {Tanaka},
  {Thayer}, {Thayer}, {Thompson}, {Tibaldo}, {Tibolla}, {Torres}, {Tosti},
  {Tramacere}, {Uchiyama}, {Usher}, {Vasileiou}, {Vilchez}, {Vitale}, {Waite},
  {Wang}, {Winer}, {Wood}, {Ylinen}, {Ziegler}, {Fermi/LAT Collaboration},
  {Ghisellini}, {Maraschi}, \& {Tavecchio}}]{2009ApJ...707L.142A}
{Abdo}, A.~A., {Ackermann}, M., {Ajello}, M., {et~al.} 2009, \apjl, 707, L142

\bibitem[{{Abdollahi} {et~al.}(2022){Abdollahi}, {Acero}, {Baldini}, {Ballet},
  {Bastieri}, {Bellazzini}, {Berenji}, {Berretta}, {Bissaldi}, {Blandford},
  {Bloom}, {Bonino}, {Brill}, {Britto}, {Bruel}, {Burnett}, {Buson}, {Cameron},
  {Caputo}, {Caraveo}, {Castro}, {Chaty}, {Cheung}, {Chiaro}, {Cibrario},
  {Ciprini}, {Coronado-Bl{\'a}zquez}, {Crnogorcevic}, {Cutini}, {D'Ammando},
  {De Gaetano}, {Digel}, {Di Lalla}, {Dirirsa}, {Di Venere}, {Dom{\'\i}nguez},
  {Fallah Ramazani}, {Fegan}, {Ferrara}, {Fiori}, {Fleischhack}, {Franckowiak},
  {Fukazawa}, {Funk}, {Fusco}, {Galanti}, {Gammaldi}, {Gargano}, {Garrappa},
  {Gasparrini}, {Giacchino}, {Giglietto}, {Giordano}, {Giroletti}, {Glanzman},
  {Green}, {Grenier}, {Grondin}, {Guillemot}, {Guiriec}, {Gustafsson},
  {Harding}, {Hays}, {Hewitt}, {Horan}, {Hou}, {J{\'o}hannesson}, {Karwin},
  {Kayanoki}, {Kerr}, {Kuss}, {Landriu}, {Larsson}, {Latronico},
  {Lemoine-Goumard}, {Li}, {Liodakis}, {Longo}, {Loparco}, {Lott}, {Lubrano},
  {Maldera}, {Malyshev}, {Manfreda}, {Mart{\'\i}-Devesa}, {Mazziotta}, {Mereu},
  {Meyer}, {Michelson}, {Mirabal}, {Mitthumsiri}, {Mizuno}, {Moiseev},
  {Monzani}, {Morselli}, {Moskalenko}, {Negro}, {Nuss}, {Omodei}, {Orienti},
  {Orlando}, {Paneque}, {Pei}, {Perkins}, {Persic}, {Pesce-Rollins},
  {Petrosian}, {Pillera}, {Poon}, {Porter}, {Principe}, {Rain{\`o}}, {Rando},
  {Rani}, {Razzano}, {Razzaque}, {Reimer}, {Reimer}, {Reposeur},
  {S{\'a}nchez-Conde}, {Saz Parkinson}, {Scotton}, {Serini}, {Sgr{\`o}},
  {Siskind}, {Smith}, {Spandre}, {Spinelli}, {Sueoka}, {Suson}, {Tajima},
  {Tak}, {Thayer}, {Thompson}, {Torres}, {Troja}, {Valverde}, {Wood}, \&
  {Zaharijas}}]{2022ApJS..260...53A}
{Abdollahi}, S., {Acero}, F., {Baldini}, L., {et~al.} 2022, \apjs, 260, 53

\bibitem[{{Ballet} {et~al.}(2026){Ballet}, {Bruel}, {Burnett}, \&
  {Lott}}]{2026arXiv260222148B}
{Ballet}, J., {Bruel}, P., {Burnett}, T.~H., \& {Lott}, B. 2026, arXiv
  e-prints, arXiv:2602.22148

\bibitem[{{Berton} {et~al.}(2019){Berton}, {Congiu}, {Ciroi}, {Komossa},
  {Frezzato}, {Di Mille}, {Ant{\'o}n}, {Antonucci}, {Caccianiga}, {Coppi},
  {J{\"a}rvel{\"a}}, {Kotilainen}, {L{\"a}hteenm{\"a}ki}, {Mathur}, {Chen},
  {Cracco}, {La Mura}, \& {Rafanelli}}]{2019AJ....157...48B}
{Berton}, M., {Congiu}, E., {Ciroi}, S., {et~al.} 2019, \aj, 157, 48

\bibitem[{{Boller} {et~al.}(1996){Boller}, {Brandt}, \&
  {Fink}}]{1996A&A...305...53B}
{Boller}, T., {Brandt}, W.~N., \& {Fink}, H. 1996, \aap, 305, 53

\bibitem[{{Bonson} {et~al.}(2018){Bonson}, {Gallo}, {Wilkins}, \&
  {Fabian}}]{2018MNRAS.477.3247B}
{Bonson}, K., {Gallo}, L.~C., {Wilkins}, D.~R., \& {Fabian}, A.~C. 2018,
  \mnras, 477, 3247

\bibitem[{{Boroson}(2002)}]{2002ApJ...565...78B}
{Boroson}, T.~A. 2002, \apj, 565, 78

\bibitem[{{Boroson} \& {Green}(1992)}]{1992ApJS...80..109B}
{Boroson}, T.~A., \& {Green}, R.~F. 1992, \apjs, 80, 109

\bibitem[{{Calderone} {et~al.}(2011){Calderone}, {Foschini}, {Ghisellini},
  {Colpi}, {Maraschi}, {Tavecchio}, {Decarli}, \&
  {Tagliaferri}}]{2011MNRAS.413.2365C}
{Calderone}, G., {Foschini}, L., {Ghisellini}, G., {et~al.} 2011, \mnras, 413,
  2365

\bibitem[{{Chen} {et~al.}(2024){Chen}, {Kharb}, {Silpa}, {Nandi}, {Berton},
  {J{\"a}rvel{\"a}}, {Laor}, {Behar}, {Foschini}, {Vietri}, {Gu}, {La Mura},
  {Crepaldi}, \& {Zhou}}]{2024ApJ...963...32C}
{Chen}, S., {Kharb}, P., {Silpa}, S., {et~al.} 2024, \apj, 963, 32

\bibitem[{{Collin} \& {Kawaguchi}(2004)}]{2004A&A...426..797C}
{Collin}, S., \& {Kawaguchi}, T. 2004, \aap, 426, 797

\bibitem[{{Crenshaw} {et~al.}(2003){Crenshaw}, {Kraemer}, \&
  {Gabel}}]{2003AJ....126.1690C}
{Crenshaw}, D.~M., {Kraemer}, S.~B., \& {Gabel}, J.~R. 2003, \aj, 126, 1690

\bibitem[{{Decarli} {et~al.}(2008){Decarli}, {Dotti}, {Fontana}, \&
  {Haardt}}]{2008MNRAS.386L..15D}
{Decarli}, R., {Dotti}, M., {Fontana}, M., \& {Haardt}, F. 2008, \mnras, 386,
  L15

\bibitem[{{DESI Collaboration} {et~al.}(2025){DESI Collaboration},
  {Abdul-Karim}, {Adame}, {Aguado}, \& collaboration}]{2025arXiv250314745D}
{DESI Collaboration}, {Abdul-Karim}, M., {Adame}, A.~G., {Aguado}, D., \&
  collaboration, D. 2025, arXiv e-prints, arXiv:2503.14745

\bibitem[{{D'Onofrio} {et~al.}(2024){D'Onofrio}, {Marziani}, {Chiosi}, \&
  {Negrete}}]{2024Univ...10..254D}
{D'Onofrio}, M., {Marziani}, P., {Chiosi}, C., \& {Negrete}, C.~A. 2024,
  Universe, 10, 254

\bibitem[{{Du} \& {Wang}(2019)}]{2019ApJ...886...42D}
{Du}, P., \& {Wang}, J.-M. 2019, \apj, 886, 42

\bibitem[{{Evans} {et~al.}(2024){Evans}, {Evans}, {Mart{\'\i}nez-Galarza},
  {Miller}, {Primini}, {Azadi}, {Burke}, {Civano}, {D'Abrusco}, {Fabbiano},
  {Graessle}, {Grier}, {Houck}, {Lauer}, {McCollough}, {Nowak}, {Plummer},
  {Rots}, {Siemiginowska}, \& {Tibbetts}}]{2024ApJS..274...22E}
{Evans}, I.~N., {Evans}, J.~D., {Mart{\'\i}nez-Galarza}, J.~R., {et~al.} 2024,
  \apjs, 274, 22

\bibitem[{{Evans} {et~al.}(2023){Evans}, {Page}, {Beardmore}, {Eyles-Ferris},
  {Osborne}, {Campana}, {Kennea}, \& {Cenko}}]{2023MNRAS.518..174E}
{Evans}, P.~A., {Page}, K.~L., {Beardmore}, A.~P., {et~al.} 2023, \mnras, 518,
  174

\bibitem[{{Fabian}(2012)}]{2012ARA&A..50..455F}
{Fabian}, A.~C. 2012, \araa, 50, 455

\bibitem[{{Fabian} {et~al.}(2015){Fabian}, {Lohfink}, {Kara}, {Parker},
  {Vasudevan}, \& {Reynolds}}]{2015MNRAS.451.4375F}
{Fabian}, A.~C., {Lohfink}, A., {Kara}, E., {et~al.} 2015, \mnras, 451, 4375

\bibitem[{{Fabian} {et~al.}(2009){Fabian}, {Zoghbi}, {Ross}, {Uttley}, {Gallo},
  {Brandt}, {Blustin}, {Boller}, {Caballero-Garcia}, {Larsson}, {Miller},
  {Miniutti}, {Ponti}, {Reis}, {Reynolds}, {Tanaka}, \&
  {Young}}]{2009Natur.459..540F}
{Fabian}, A.~C., {Zoghbi}, A., {Ross}, R.~R., {et~al.} 2009, \nat, 459, 540

\bibitem[{{Ferland} {et~al.}(2020){Ferland}, {Done}, {Jin}, {Landt}, \&
  {Ward}}]{2020MNRAS.494.5917F}
{Ferland}, G.~J., {Done}, C., {Jin}, C., {Landt}, H., \& {Ward}, M.~J. 2020,
  \mnras, 494, 5917

\bibitem[{{Foschini} {et~al.}(2015){Foschini}, {Berton}, {Caccianiga}, {Ciroi},
  {Cracco}, {Peterson}, {Angelakis}, {Braito}, {Fuhrmann}, {Gallo}, {Grupe},
  {J{\"a}rvel{\"a}}, {Kaufmann}, {Komossa}, {Kovalev}, {L{\"a}hteenm{\"a}ki},
  {Lisakov}, {Lister}, {Mathur}, {Richards}, {Romano}, {Sievers},
  {Tagliaferri}, {Tammi}, {Tibolla}, {Tornikoski}, {Vercellone}, {La Mura},
  {Maraschi}, \& {Rafanelli}}]{2015ANA...575A..13F}
{Foschini}, L., {Berton}, M., {Caccianiga}, A., {et~al.} 2015, \aap, 575, A13

\bibitem[{{Frederick} {et~al.}(2021){Frederick}, {Gezari}, {Graham},
  {Sollerman}, {van Velzen}, {Perley}, {Stern}, {Ward}, {Hammerstein}, {Hung},
  {Yan}, {Andreoni}, {Bellm}, {Duev}, {Kowalski}, {Mahabal}, {Masci},
  {Medford}, {Rusholme}, {Smith}, \& {Walters}}]{2021ApJ...920...56F}
{Frederick}, S., {Gezari}, S., {Graham}, M.~J., {et~al.} 2021, \apj, 920, 56

\bibitem[{{Fuhrmann} {et~al.}(2016){Fuhrmann}, {Karamanavis}, {Komossa},
  {Angelakis}, {Krichbaum}, {Schulz}, {Kreikenbohm}, {Kadler}, {Myserlis},
  {Ros}, {Nestoras}, \& {Zensus}}]{2016RAA....16..176F}
{Fuhrmann}, L., {Karamanavis}, V., {Komossa}, S., {et~al.} 2016, Research in
  Astronomy and Astrophysics, 16, 176

\bibitem[{{Goad} {et~al.}(2012){Goad}, {Korista}, \&
  {Ruff}}]{2012MNRAS.426.3086G}
{Goad}, M.~R., {Korista}, K.~T., \& {Ruff}, A.~J. 2012, \mnras, 426, 3086

\bibitem[{{Goodrich}(1989)}]{1989ApJ...342..224G}
{Goodrich}, R.~W. 1989, \apj, 342, 224

\bibitem[{{Gordon} {et~al.}(2021){Gordon}, {Boyce}, {O'Dea}, {Rudnick},
  {Andernach}, {Vantyghem}, {Baum}, {Bui}, {Dionyssiou}, {Safi-Harb}, \&
  {Sander}}]{2021ApJS..255...30G}
{Gordon}, Y.~A., {Boyce}, M.~M., {O'Dea}, C.~P., {et~al.} 2021, \apjs, 255, 30

\bibitem[{{Greene} \& {Ho}(2005)}]{2005ApJ...630..122G}
{Greene}, J.~E., \& {Ho}, L.~C. 2005, \apj, 630, 122

\bibitem[{{Grupe}(2004)}]{2004AJ....127.1799G}
{Grupe}, D. 2004, \aj, 127, 1799

\bibitem[{{Grupe} {et~al.}(2010){Grupe}, {Komossa}, {Leighly}, \&
  {Page}}]{2010ApJS..187...64G}
{Grupe}, D., {Komossa}, S., {Leighly}, K.~M., \& {Page}, K.~L. 2010, \apjs,
  187, 64

\bibitem[{{Grupe} \& {Mathur}(2004)}]{2004ApJ...606L..41G}
{Grupe}, D., \& {Mathur}, S. 2004, \apjl, 606, L41

\bibitem[{{Guo} {et~al.}(2018){Guo}, {Shen}, \& {Wang}}]{2018ascl.soft09008G}
{Guo}, H., {Shen}, Y., \& {Wang}, S. 2018, {PyQSOFit: Python code to fit the
  spectrum of quasars}, , , ascl:1809.008

\bibitem[{{Hopkins} \& {Elvis}(2010)}]{2010MNRAS.401....7H}
{Hopkins}, P.~F., \& {Elvis}, M. 2010, \mnras, 401, 7

\bibitem[{{Hopkins} {et~al.}(2016){Hopkins}, {Torrey}, {Faucher-Gigu{\`e}re},
  {Quataert}, \& {Murray}}]{2016MNRAS.458..816H}
{Hopkins}, P.~F., {Torrey}, P., {Faucher-Gigu{\`e}re}, C.-A., {Quataert}, E.,
  \& {Murray}, N. 2016, \mnras, 458, 816

\bibitem[{{Ili{\'c}} {et~al.}(2023){Ili{\'c}}, {Raki{\'c}}, \&
  {Popovi{\'c}}}]{2023ApJS..267...19I}
{Ili{\'c}}, D., {Raki{\'c}}, N., \& {Popovi{\'c}}, L.~{\v{C}}. 2023, \apjs,
  267, 19

\bibitem[{{Jordi} {et~al.}(2006){Jordi}, {Grebel}, \&
  {Ammon}}]{2006A&A...460..339J}
{Jordi}, K., {Grebel}, E.~K., \& {Ammon}, K. 2006, \aap, 460, 339

\bibitem[{{Kara} {et~al.}(2017){Kara}, {Garc{\'\i}a}, {Lohfink}, {Fabian},
  {Reynolds}, {Tombesi}, \& {Wilkins}}]{2017MNRAS.468.3489K}
{Kara}, E., {Garc{\'\i}a}, J.~A., {Lohfink}, A., {et~al.} 2017, \mnras, 468,
  3489

\bibitem[{{Kellermann} {et~al.}(1989){Kellermann}, {Sramek}, {Schmidt},
  {Shaffer}, \& {Green}}]{1989AJ.....98.1195K}
{Kellermann}, K.~I., {Sramek}, R., {Schmidt}, M., {Shaffer}, D.~B., \& {Green},
  R. 1989, \aj, 98, 1195

\bibitem[{{Kewley} {et~al.}(2001){Kewley}, {Dopita}, {Sutherland}, {Heisler},
  \& {Trevena}}]{2001ApJ...556..121K}
{Kewley}, L.~J., {Dopita}, M.~A., {Sutherland}, R.~S., {Heisler}, C.~A., \&
  {Trevena}, J. 2001, \apj, 556, 121

\bibitem[{{Komossa} {et~al.}(2006){Komossa}, {Voges}, {Xu}, {Mathur}, {Adorf},
  {Lemson}, {Duschl}, \& {Grupe}}]{2006AJ....132..531K}
{Komossa}, S., {Voges}, W., {Xu}, D., {et~al.} 2006, \aj, 132, 531

\bibitem[{{Komossa} {et~al.}(2008){Komossa}, {Xu}, {Zhou}, {Storchi-Bergmann},
  \& {Binette}}]{2008ApJ...680..926K}
{Komossa}, S., {Xu}, D., {Zhou}, H., {Storchi-Bergmann}, T., \& {Binette}, L.
  2008, \apj, 680, 926

\bibitem[{{Komossa} {et~al.}(2018){Komossa}, {Xu}, \&
  {Wagner}}]{2018MNRAS.477.5115K}
{Komossa}, S., {Xu}, D.~W., \& {Wagner}, A.~Y. 2018, \mnras, 477, 5115

\bibitem[{{Koz{\'a}k} {et~al.}(2024){Koz{\'a}k}, {Frey}, \&
  {Gab{\'a}nyi}}]{2024Galax..12....8K}
{Koz{\'a}k}, B., {Frey}, S., \& {Gab{\'a}nyi}, K.~{\'E}. 2024, Galaxies, 12, 8

\bibitem[{{Kshama} {et~al.}(2017){Kshama}, {Paliya}, \&
  {Stalin}}]{2017MNRAS.466.2679K}
{Kshama}, S.~K., {Paliya}, V.~S., \& {Stalin}, C.~S. 2017, \mnras, 466, 2679

\bibitem[{{Lacy} {et~al.}(2020){Lacy}, {Baum}, {Chandler}, {Chatterjee},
  {Clarke}, {Deustua}, {English}, {Farnes}, {Gaensler}, {Gugliucci},
  {Hallinan}, {Kent}, {Kimball}, {Law}, {Lazio}, {Marvil}, {Mao}, {Medlin},
  {Mooley}, {Murphy}, {Myers}, {Osten}, {Richards}, {Rosolowsky}, {Rudnick},
  {Schinzel}, {Sivakoff}, {Sjouwerman}, {Taylor}, {White}, {Wrobel},
  {Andernach}, {Beasley}, {Berger}, {Bhatnager}, {Birkinshaw}, {Bower},
  {Brandt}, {Brown}, {Burke-Spolaor}, {Butler}, {Comerford}, {Demorest}, {Fu},
  {Giacintucci}, {Golap}, {G{\"u}th}, {Hales}, {Hiriart}, {Hodge}, {Horesh},
  {Ivezi{\'c}}, {Jarvis}, {Kamble}, {Kassim}, {Liu}, {Loinard}, {Lyons},
  {Masters}, {Mezcua}, {Moellenbrock}, {Mroczkowski}, {Nyland}, {O'Dea},
  {O'Sullivan}, {Peters}, {Radford}, {Rao}, {Robnett}, {Salcido}, {Shen},
  {Sobotka}, {Witz}, {Vaccari}, {van Weeren}, {Vargas}, {Williams}, \&
  {Yoon}}]{2020PASP..132c5001L}
{Lacy}, M., {Baum}, S.~A., {Chandler}, C.~J., {et~al.} 2020, \pasp, 132, 035001

\bibitem[{{Lang} {et~al.}(2016){Lang}, {Hogg}, \&
  {Mykytyn}}]{2016ascl.soft04008L}
{Lang}, D., {Hogg}, D.~W., \& {Mykytyn}, D. 2016, {The Tractor: Probabilistic
  astronomical source detection and measurement}, Astrophysics Source Code
  Library, record ascl:1604.008, , , ascl:1604.008

\bibitem[{{Leighly}(1999{\natexlab{a}})}]{1999ApJS..125..297L}
{Leighly}, K.~M. 1999{\natexlab{a}}, \apjs, 125, 297

\bibitem[{{Leighly}(1999{\natexlab{b}})}]{1999ApJS..125..317L}
---. 1999{\natexlab{b}}, \apjs, 125, 317

\bibitem[{{Lu} \& {Yu}(2001)}]{2001MNRAS.324..653L}
{Lu}, Y., \& {Yu}, Q. 2001, \mnras, 324, 653

\bibitem[{{Marziani} {et~al.}(2018){Marziani}, {Dultzin}, {Sulentic}, {Del
  Olmo}, {Negrete}, {Mart{\'\i}nez-Aldama}, {D'Onofrio}, {Bon}, {Bon}, \&
  {Stirpe}}]{2018FrASS...5....6M}
{Marziani}, P., {Dultzin}, D., {Sulentic}, J.~W., {et~al.} 2018, Frontiers in
  Astronomy and Space Sciences, 5, 6

\bibitem[{{Mathur}(2000)}]{2000MNRAS.314L..17M}
{Mathur}, S. 2000, \mnras, 314, L17

\bibitem[{{McLure} \& {Dunlop}(2004)}]{2004MNRAS.352.1390M}
{McLure}, R.~J., \& {Dunlop}, J.~S. 2004, \mnras, 352, 1390

\bibitem[{{Mej{\'\i}a-Restrepo} {et~al.}(2018){Mej{\'\i}a-Restrepo}, {Lira},
  {Netzer}, {Trakhtenbrot}, \& {Capellupo}}]{2018NatAs...2...63M}
{Mej{\'\i}a-Restrepo}, J.~E., {Lira}, P., {Netzer}, H., {Trakhtenbrot}, B., \&
  {Capellupo}, D.~M. 2018, Nature Astronomy, 2, 63

\bibitem[{{Merloni} {et~al.}(2024){Merloni}, {Lamer}, {Liu}, {Ramos-Ceja},
  {Brunner}, {Bulbul}, {Dennerl}, {Doroshenko}, {Freyberg}, {Friedrich},
  {Gatuzz}, {Georgakakis}, {Haberl}, {Igo}, {Kreykenbohm}, {Liu}, {Maitra},
  {Malyali}, {Mayer}, {Nandra}, {Predehl}, {Robrade}, {Salvato}, {Sanders},
  {Stewart}, {Tub{\'\i}n-Arenas}, {Weber}, {Wilms}, {Arcodia}, {Artis},
  {Aschersleben}, {Avakyan}, {Aydar}, {Bahar}, {Balzer}, {Becker}, {Berger},
  {Boller}, {Bornemann}, {Br{\"u}ggen}, {Brusa}, {Buchner}, {Burwitz},
  {Camilloni}, {Clerc}, {Comparat}, {Coutinho}, {Czesla}, {Dannhauer},
  {Dauner}, {Dauser}, {Dietl}, {Dolag}, {Dwelly}, {Egg}, {Ehl}, {Freund},
  {Friedrich}, {Gaida}, {Garrel}, {Ghirardini}, {Gokus}, {Gr{\"u}nwald},
  {Grandis}, {Grotova}, {Gruen}, {Gueguen}, {H{\"a}mmerich}, {Hamaus},
  {Hasinger}, {Haubner}, {Homan}, {Ider Chitham}, {Joseph}, {Joyce},
  {K{\"o}nig}, {Kaltenbrunner}, {Khokhriakova}, {Kink}, {Kirsch}, {Kluge},
  {Knies}, {Krippendorf}, {Krumpe}, {Kurpas}, {Li}, {Liu}, {Locatelli},
  {Lorenz}, {M{\"u}ller}, {Magaudda}, {Mannes}, {McCall}, {Meidinger},
  {Michailidis}, {Migkas}, {Mu{\~n}oz-Giraldo}, {Musiimenta}, {Nguyen-Dang},
  {Ni}, {Olechowska}, {Ota}, {Pacaud}, {Pasini}, {Perinati}, {Pires},
  {Pommranz}, {Ponti}, {Poppenhaeger}, {P{\"u}hlhofer}, {Rau}, {Reh},
  {Reiprich}, {Roster}, {Saeedi}, {Santangelo}, {Sasaki}, {Schmitt},
  {Schneider}, {Schrabback}, {Schuster}, {Schwope}, {Seppi}, {Serim},
  {Shreeram}, {Sokolova-Lapa}, {Starck}, {Stelzer}, {Stierhof}, {Suleimanov},
  {Tenzer}, {Traulsen}, {Tr{\"u}mper}, {Tsuge}, {Urrutia}, {Veronica},
  {Waddell}, {Willer}, {Wolf}, {Yeung}, {Zainab}, {Zangrandi}, {Zhang},
  {Zhang}, \& {Zheng}}]{2024A&A...682A..34M}
{Merloni}, A., {Lamer}, G., {Liu}, T., {et~al.} 2024, \aap, 682, A34

\bibitem[{{Moran} {et~al.}(1996){Moran}, {Halpern}, \&
  {Helfand}}]{1996ApJS..106..341M}
{Moran}, E.~C., {Halpern}, J.~P., \& {Helfand}, D.~J. 1996, \apjs, 106, 341

\bibitem[{{Mundo} {et~al.}(2020){Mundo}, {Kara}, {Cackett}, {Fabian}, {Jiang},
  {Mushotzky}, {Parker}, {Pinto}, {Reynolds}, \&
  {Zoghbi}}]{2020MNRAS.496.2922M}
{Mundo}, S.~A., {Kara}, E., {Cackett}, E.~M., {et~al.} 2020, \mnras, 496, 2922

\bibitem[{{Netzer}(2019)}]{2019MNRAS.488.5185N}
{Netzer}, H. 2019, \mnras, 488, 5185

\bibitem[{{Olgu{\'\i}n-Iglesias} {et~al.}(2016){Olgu{\'\i}n-Iglesias},
  {Le{\'o}n-Tavares}, {Kotilainen}, {Chavushyan}, {Tornikoski}, {Valtaoja},
  {A{\~n}orve}, {Vald{\'e}s}, \& {Carrasco}}]{2016MNRAS.460.3202O}
{Olgu{\'\i}n-Iglesias}, A., {Le{\'o}n-Tavares}, J., {Kotilainen}, J.~K.,
  {et~al.} 2016, \mnras, 460, 3202

\bibitem[{{Osterbrock} \& {Pogge}(1985)}]{1985ApJ...297..166O}
{Osterbrock}, D.~E., \& {Pogge}, R.~W. 1985, \apj, 297, 166

\bibitem[{{Paliya}(2019)}]{2019JApA...40...39P}
{Paliya}, V.~S. 2019, Journal of Astrophysics and Astronomy, 40, 39

\bibitem[{{Paliya} {et~al.}(2018){Paliya}, {Ajello}, {Rakshit}, {Mandal},
  {Stalin}, {Kaur}, \& {Hartmann}}]{2018ApJ...853L...2P}
{Paliya}, V.~S., {Ajello}, M., {Rakshit}, S., {et~al.} 2018, \apjl, 853, L2

\bibitem[{{Paliya} {et~al.}(2019){Paliya}, {Parker}, {Jiang}, {Fabian},
  {Brenneman}, {Ajello}, \& {Hartmann}}]{2019ApJ...872..169P}
{Paliya}, V.~S., {Parker}, M.~L., {Jiang}, J., {et~al.} 2019, \apj, 872, 169

\bibitem[{{Paliya} {et~al.}(2024){Paliya}, {Stalin}, {Dom{\'\i}nguez}, \&
  {Saikia}}]{2024MNRAS.527.7055P}
{Paliya}, V.~S., {Stalin}, C.~S., {Dom{\'\i}nguez}, A., \& {Saikia}, D.~J.
  2024, \mnras, 527, 7055

\bibitem[{{Paliya} {et~al.}(2013){Paliya}, {Stalin}, {Kumar}, {Kumar}, {Bhatt},
  {Pandey}, \& {Yadav}}]{2013MNRAS.428.2450P}
{Paliya}, V.~S., {Stalin}, C.~S., {Kumar}, B., {et~al.} 2013, \mnras, 428, 2450

\bibitem[{{Paliya} {et~al.}(2015){Paliya}, {Stalin}, \&
  {Ravikumar}}]{2015AJ....149...41P}
{Paliya}, V.~S., {Stalin}, C.~S., \& {Ravikumar}, C.~D. 2015, \aj, 149, 41

\bibitem[{{Paliya} {et~al.}(2020){Paliya}, {P{\'e}rez}, {Garc{\'\i}a-Benito},
  {Ajello}, {Prada}, {Alberdi}, {Suh}, {Chandra}, {Dom{\'\i}nguez}, {Marchesi},
  {Di Matteo}, {Hartmann}, \& {Chiaberge}}]{2020ApJ...892..133P}
{Paliya}, V.~S., {P{\'e}rez}, E., {Garc{\'\i}a-Benito}, R., {et~al.} 2020,
  \apj, 892, 133

\bibitem[{{Panda} {et~al.}(2019){Panda}, {Marziani}, \&
  {Czerny}}]{2019ApJ...882...79P}
{Panda}, S., {Marziani}, P., \& {Czerny}, B. 2019, \apj, 882, 79

\bibitem[{{Parker} {et~al.}(2014){Parker}, {Wilkins}, {Fabian}, {Grupe},
  {Dauser}, {Matt}, {Harrison}, {Brenneman}, {Boggs}, {Christensen}, {Craig},
  {Gallo}, {Hailey}, {Kara}, {Komossa}, {Marinucci}, {Miller}, {Risaliti},
  {Stern}, {Walton}, \& {Zhang}}]{2014MNRAS.443.1723P}
{Parker}, M.~L., {Wilkins}, D.~R., {Fabian}, A.~C., {et~al.} 2014, \mnras, 443,
  1723

\bibitem[{{Parker} {et~al.}(2019){Parker}, {Longinotti}, {Schartel}, {Grupe},
  {Komossa}, {Kriss}, {Fabian}, {Gallo}, {Harrison}, {Jiang}, {Kara},
  {Krongold}, {Matzeu}, {Pinto}, \& {Santos-Lle{\'o}}}]{2019MNRAS.490..683P}
{Parker}, M.~L., {Longinotti}, A.~L., {Schartel}, N., {et~al.} 2019, \mnras,
  490, 683

\bibitem[{{Ponti} {et~al.}(2012){Ponti}, {Papadakis}, {Bianchi}, {Guainazzi},
  {Matt}, {Uttley}, \& {Bonilla}}]{2012A&A...542A..83P}
{Ponti}, G., {Papadakis}, I., {Bianchi}, S., {et~al.} 2012, \aap, 542, A83

\bibitem[{{Rakshit} {et~al.}(2019){Rakshit}, {Johnson}, {Stalin}, {Gandhi}, \&
  {Hoenig}}]{2019MNRAS.483.2362R}
{Rakshit}, S., {Johnson}, A., {Stalin}, C.~S., {Gandhi}, P., \& {Hoenig}, S.
  2019, \mnras, 483, 2362

\bibitem[{{Rakshit} \& {Stalin}(2017)}]{2017ApJ...842...96R}
{Rakshit}, S., \& {Stalin}, C.~S. 2017, \apj, 842, 96

\bibitem[{{Rakshit} {et~al.}(2017){Rakshit}, {Stalin}, {Chand}, \&
  {Zhang}}]{2017ApJS..229...39R}
{Rakshit}, S., {Stalin}, C.~S., {Chand}, H., \& {Zhang}, X.-G. 2017, \apjs,
  229, 39

\bibitem[{{Rakshit} {et~al.}(2018){Rakshit}, {Stalin}, {Hota}, \&
  {Konar}}]{2018ApJ...869..173R}
{Rakshit}, S., {Stalin}, C.~S., {Hota}, A., \& {Konar}, C. 2018, \apj, 869, 173

\bibitem[{{Rakshit} \& {Woo}(2018)}]{2018ApJ...865....5R}
{Rakshit}, S., \& {Woo}, J.-H. 2018, \apj, 865, 5

\bibitem[{{Ren} {et~al.}(2024){Ren}, {Guo}, {Shen}, {Silverman}, {Burke},
  {Wang}, \& {Wang}}]{2024ApJ...974..153R}
{Ren}, W., {Guo}, H., {Shen}, Y., {et~al.} 2024, \apj, 974, 153

\bibitem[{{Richards} {et~al.}(2006){Richards}, {Lacy}, {Storrie-Lombardi},
  {Hall}, {Gallagher}, {Hines}, {Fan}, {Papovich}, {Vanden Berk}, {Trammell},
  {Schneider}, {Vestergaard}, {York}, {Jester}, {Anderson}, {Budav{\'a}ri}, \&
  {Szalay}}]{2006ApJS..166..470R}
{Richards}, G.~T., {Lacy}, M., {Storrie-Lombardi}, L.~J., {et~al.} 2006, \apjs,
  166, 470

\bibitem[{{Sabater} {et~al.}(2019){Sabater}, {Best}, {Hardcastle}, {Shimwell},
  {Tasse}, {Williams}, {Br{\"u}ggen}, {Cochrane}, {Croston}, {de Gasperin},
  {Duncan}, {G{\"u}rkan}, {Mechev}, {Morabito}, {Prandoni}, {R{\"o}ttgering},
  {Smith}, {Harwood}, {Mingo}, {Mooney}, \& {Saxena}}]{2019A&A...622A..17S}
{Sabater}, J., {Best}, P.~N., {Hardcastle}, M.~J., {et~al.} 2019, \aap, 622,
  A17

\bibitem[{{Savaglio} {et~al.}(2005){Savaglio}, {Glazebrook}, {Le Borgne},
  {Juneau}, {Abraham}, {Chen}, {Crampton}, {McCarthy}, {Carlberg}, {Marzke},
  {Roth}, {J{\o}rgensen}, \& {Murowinski}}]{2005ApJ...635..260S}
{Savaglio}, S., {Glazebrook}, K., {Le Borgne}, D., {et~al.} 2005, \apj, 635,
  260

\bibitem[{{Schlegel} {et~al.}(1998){Schlegel}, {Finkbeiner}, \&
  {Davis}}]{1998ApJ...500..525S}
{Schlegel}, D.~J., {Finkbeiner}, D.~P., \& {Davis}, M. 1998, \apj, 500, 525

\bibitem[{{Shen}(2013)}]{2013BASI...41...61S}
{Shen}, Y. 2013, Bulletin of the Astronomical Society of India, 41, 61

\bibitem[{{Shen} {et~al.}(2011){Shen}, {Richards}, {Strauss}, {Hall},
  {Schneider}, {Snedden}, {Bizyaev}, {Brewington}, {Malanushenko},
  {Malanushenko}, {Oravetz}, {Pan}, \& {Simmons}}]{2011ApJS..194...45S}
{Shen}, Y., {Richards}, G.~T., {Strauss}, M.~A., {et~al.} 2011, \apjs, 194, 45

\bibitem[{{Shuder} \& {Osterbrock}(1981)}]{1981ApJ...250...55S}
{Shuder}, J.~M., \& {Osterbrock}, D.~E. 1981, \apj, 250, 55

\bibitem[{{Singh} \& {Chand}(2018)}]{2018MNRAS.480.1796S}
{Singh}, V., \& {Chand}, H. 2018, \mnras, 480, 1796

\bibitem[{{Sudan} {et~al.}(2025){Sudan}, {Chand}, {Wiita}, \&
  {Kumar}}]{2025MNRAS.543..121S}
{Sudan}, M., {Chand}, H., {Wiita}, P.~J., \& {Kumar}, R. 2025, \mnras, 543, 121

\bibitem[{{Sulentic} {et~al.}(2000){Sulentic}, {Marziani}, \&
  {Dultzin-Hacyan}}]{2000ARA&A..38..521S}
{Sulentic}, J.~W., {Marziani}, P., \& {Dultzin-Hacyan}, D. 2000, \araa, 38, 521

\bibitem[{{Sulentic} {et~al.}(2002){Sulentic}, {Marziani}, {Zamanov}, {Bachev},
  {Calvani}, \& {Dultzin-Hacyan}}]{2002ApJ...566L..71S}
{Sulentic}, J.~W., {Marziani}, P., {Zamanov}, R., {et~al.} 2002, \apjl, 566,
  L71

\bibitem[{{Umayal} {et~al.}(2025){Umayal}, {Paliya}, {Saikia}, {Stalin},
  {Muneer}, \& {Gopinathan}}]{2025ApJ...995..125U}
{Umayal}, S., {Paliya}, V.~S., {Saikia}, D.~J., {et~al.} 2025, \apj, 995, 125

\bibitem[{{Vanden Berk} {et~al.}(2006){Vanden Berk}, {Shen}, {Yip},
  {Schneider}, {Connolly}, {Burton}, {Jester}, {Hall}, {Szalay}, \&
  {Brinkmann}}]{2006AJ....131...84V}
{Vanden Berk}, D.~E., {Shen}, J., {Yip}, C.-W., {et~al.} 2006, \aj, 131, 84

\bibitem[{{Veilleux} \& {Osterbrock}(1987)}]{1987ApJS...63..295V}
{Veilleux}, S., \& {Osterbrock}, D.~E. 1987, \apjs, 63, 295

\bibitem[{{V{\'e}ron-Cetty} \& {V{\'e}ron}(2010)}]{2010A&A...518A..10V}
{V{\'e}ron-Cetty}, M.-P., \& {V{\'e}ron}, P. 2010, \aap, 518, A10

\bibitem[{{V{\'e}ron-Cetty} {et~al.}(2001){V{\'e}ron-Cetty}, {V{\'e}ron}, \&
  {Gon{\c{c}}alves}}]{2001A&A...372..730V}
{V{\'e}ron-Cetty}, M.-P., {V{\'e}ron}, P., \& {Gon{\c{c}}alves}, A.~C. 2001,
  \aap, 372, 730

\bibitem[{{Vestergaard} \& {Peterson}(2006)}]{2006ApJ...641..689V}
{Vestergaard}, M., \& {Peterson}, B.~M. 2006, \apj, 641, 689

\bibitem[{{Webb} {et~al.}(2020){Webb}, {Coriat}, {Traulsen}, {Ballet}, {Motch},
  {Carrera}, {Koliopanos}, {Authier}, {de la Calle}, {Ceballos}, {Colomo},
  {Chuard}, {Freyberg}, {Garcia}, {Kolehmainen}, {Lamer}, {Lin}, {Maggi},
  {Michel}, {Page}, {Page}, {Perea-Calderon}, {Pineau}, {Rodriguez}, {Rosen},
  {Santos Lleo}, {Saxton}, {Schwope}, {Tom{\'a}s}, {Watson}, \&
  {Zakardjian}}]{2020A&A...641A.136W}
{Webb}, N.~A., {Coriat}, M., {Traulsen}, I., {et~al.} 2020, \aap, 641, A136

\bibitem[{{White} {et~al.}(1997){White}, {Becker}, {Helfand}, \&
  {Gregg}}]{1997ApJ...475..479W}
{White}, R.~L., {Becker}, R.~H., {Helfand}, D.~J., \& {Gregg}, M.~D. 1997,
  \apj, 475, 479

\bibitem[{{Whittle}(1992)}]{1992ApJS...79...49W}
{Whittle}, M. 1992, \apjs, 79, 49

\bibitem[{{Williams} {et~al.}(2002){Williams}, {Pogge}, \&
  {Mathur}}]{2002AJ....124.3042W}
{Williams}, R.~J., {Pogge}, R.~W., \& {Mathur}, S. 2002, \aj, 124, 3042

\bibitem[{{Winkler}(1992)}]{1992MNRAS.257..677W}
{Winkler}, H. 1992, \mnras, 257, 677

\bibitem[{{Wright} {et~al.}(2019){Wright}, {Eisenhardt}, {Mainzer}, {Ressler},
  {Cutri}, {Jarrett}, {Kirkpatrick}, {Padgett}, {McMillan}, {Skrutskie},
  {Stanford}, {Cohen}, {Walker}, {Mather}, {Leisawitz}, {Gautier}, {McLean},
  {Benford}, {Lonsdale}, {Blain}, {Mendez}, {Irace}, {Duval}, {Liu}, {Royer},
  {Heinrichsen}, {Howard}, {Shannon}, {Kendall}, {Walsh}, {Larsen}, {Cardon},
  {Schick}, {Schwalm}, {Abid}, {Fabinsky}, {Naes}, \&
  {Tsai}}]{2019ipac.data...I1W}
{Wright}, E.~L., {Eisenhardt}, P. R.~M., {Mainzer}, A.~K., {et~al.} 2019,
  {AllWISE Source Catalog}, NASA IPAC DataSet, IRSA1, , , doi:10.26131/IRSA1

\bibitem[{{Xu} {et~al.}(2012){Xu}, {Komossa}, {Zhou}, {Lu}, {Li}, {Grupe},
  {Wang}, \& {Yuan}}]{2012AJ....143...83X}
{Xu}, D., {Komossa}, S., {Zhou}, H., {et~al.} 2012, \aj, 143, 83

\bibitem[{{Yip} {et~al.}(2004{\natexlab{a}}){Yip}, {Connolly}, {Szalay},
  {Budav{\'a}ri}, {SubbaRao}, {Frieman}, {Nichol}, {Hopkins}, {York},
  {Okamura}, {Brinkmann}, {Csabai}, {Thakar}, {Fukugita}, \&
  {Ivezi{\'c}}}]{2004AJ....128..585Y}
{Yip}, C.~W., {Connolly}, A.~J., {Szalay}, A.~S., {et~al.} 2004{\natexlab{a}},
  \aj, 128, 585

\bibitem[{{Yip} {et~al.}(2004{\natexlab{b}}){Yip}, {Connolly}, {Vanden Berk},
  {Ma}, {Frieman}, {SubbaRao}, {Szalay}, {Richards}, {Hall}, {Schneider},
  {Hopkins}, {Trump}, \& {Brinkmann}}]{2004AJ....128.2603Y}
{Yip}, C.~W., {Connolly}, A.~J., {Vanden Berk}, D.~E., {et~al.}
  2004{\natexlab{b}}, \aj, 128, 2603

\bibitem[{{Zhou} {et~al.}(2006){Zhou}, {Wang}, {Yuan}, {Lu}, {Dong}, {Wang}, \&
  {Lu}}]{2006ApJS..166..128Z}
{Zhou}, H., {Wang}, T., {Yuan}, W., {et~al.} 2006, \apjs, 166, 128

\end{thebibliography}

\end{document}